\documentclass[12pt,a4paper]{article}

\usepackage{graphicx,physics,amsmath, amsthm,amsbsy,amssymb,amsfonts,slashed,subcaption,bbm,bm,upgreek,caption,jheppub,xcolor}
\usepackage[nice]{nicefrac}
\usepackage{comment}

\newcommand{\HC}[1]{ { \color{orange} \footnotesize (\textsf{HC}) \textsf{\textsl{#1}} } }

\title{\Large{Free Probability approach to spectral and operator statistics in Rosenzweig-Porter random matrix ensembles}}

\author[a]{Viktor Jahnke\,\href{https://orcid.org/0000-0003-0204-9716}
{\includegraphics[scale=0.05]{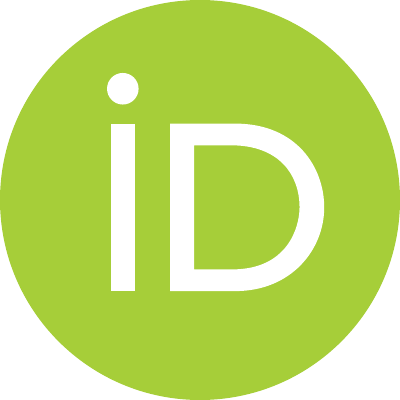}}\footnote{V.J. and P.N. contributed equally.},}
\author[b,c]{Pratik Nandy\,\href{https://orcid.org/0000-0001-5383-2458}
{\includegraphics[scale=0.05]{orcidid.pdf}},}
\author[a]{Kuntal Pal\,\href{https://orcid.org/0000-0002-1316-5213}
{\includegraphics[scale=0.05]{orcidid.pdf}},}
\author[a]{Hugo A. Camargo\,\href{https://orcid.org/0000-0002-5523-546X}
{\includegraphics[scale=0.05]{orcidid.pdf}},}
\author[a,g]{and Keun-Young Kim\,\href{https://orcid.org/0000-0002-4725-3211}
{\includegraphics[scale=0.05]{orcidid.pdf}}\,}

\emailAdd{viktorjahnke@gmail.com}
\emailAdd{pratik@yukawa.kyoto-u.ac.jp}
\emailAdd{kuntalpal@gist.ac.kr}
\emailAdd{hugocm\_89@hotmail.com}
\emailAdd{fortoe@gist.ac.kr}

\affiliation[a]{Department of Physics and Photon Science, Gwangju Institute of Science and Technology,\\
123 Cheomdan-gwagiro, Gwangju 61005, Korea}
\affiliation[b]{Center for Gravitational Physics and Quantum Information,\\ Yukawa Institute for Theoretical Physics, Kyoto University,\\ Kitashirakawa Oiwakecho, Sakyo-ku, Kyoto 606-8502, Japan}
\affiliation[c]{Division of Fundamental Mathematical Science,\\RIKEN Center for Interdisciplinary Theoretical and Mathematical Sciences (iTHEMS),\\ Wako, Saitama 351-0198, Japan}
\affiliation[g]{Research Center for Photon Science Technology, Gwangju Institute of Science and Technology, 123 Cheomdan-gwagiro, Gwangju 61005, Korea}

\abstract{Utilizing the framework of free probability, we analyze the spectral and operator statistics of the Rosenzweig--Porter random matrix ensembles, which exhibit a rich phase structure encompassing ergodic, fractal, and localized regimes. Leveraging subordination formulae, we develop a perturbative scheme that yields semi-analytic expressions for the density of states up to second order in system size, in good agreement with numerical results. We compute higher-point correlation functions in the ergodic regime using both numerical and suitable analytic approximations. Our analysis of operator statistics for various spin operators across these regimes reveals close agreement with free probability predictions in the ergodic phase, in contrast to persistent deviations observed in the fractal and localized phases, even at late times. Notably, the fractal phase exhibits \emph{partial freeness} while retaining memory of the initial spectrum, highlighting the importance of non-localized eigenstates and associated with the late-time dynamics of \emph{cumulative out-of-time-ordered-correlators} (OTOCs). Employing distance measures and statistical tools such as the $\chi^2$ statistic, Kullback--Leibler divergence, and Kolmogorov--Smirnov hypothesis testing, we define a characteristic time scale -- the \emph{free time} -- that marks the onset of the validity of free probability predictions for operator spectral statistics in the ergodic phase. Remarkably, our findings demonstrate consistency across these different approaches.}

\begin{document}

\maketitle
\flushbottom

\section{Introduction}
\label{sec:intro}

While chaos in the classical domain is a well-defined phenomenon, particularly through the understanding of initial perturbations in phase space trajectories, its extension to the quantum domain presents formidable challenges. This is largely due to the breakdown of the notion of well-defined trajectories in phase space beyond the semiclassical regime, necessitating an entirely different formulation. As such, quantum chaos has traditionally been studied through correlations among the eigenvalues of the Hamiltonian, an approach pioneered by the seminal works of Wigner, Dyson, and Mehta \cite{, Wigner1, Dyson1962a, mehta1960statistical}. These efforts sparked a wave of research, laying the foundation for studying chaotic quantum systems through statistical measures \cite{berry77, Guhr:1997ve, Oganesyan:2007wpd, Atas_rvalue, Brezin1997, Cotler:2016fpe, Gharibyan:2018jrp}.

In addition, recent advancements have shed light on quantum chaos through the study of operator growth. Here, through the Heisenberg time evolution, localized quantum information encoded in simple operators spreads throughout a system, known as information scrambling \cite{Roberts:2018mnp}. In particular, the time evolution of operators and their correlation functions \cite{Hosur:2015ylk, Cotler:2017jue}, often described in terms of diagnostic tools, such as out-of-time-order correlators (OTOCs) \cite{LarkinOTOC}, and Krylov complexity \cite{Parker:2018yvk}, have proven useful tools to diagnose quantum chaotic behavior. OTOCs, for instance, measure the sensitivity of a quantum system to initial perturbations, mirroring the classical intuition of phase space trajectories. Krylov complexity, on the other hand, traces the operator through a specific basis and recasts the problem into a particle-hopping problem in a one-dimensional chain \cite{Nandy:2024htc}. Apart from quantum many-body systems, these studies have been forefront of the holographic understanding of black holes, which are known to be maximally chaotic systems in nature \cite{Sekino:2008he, Maldacena:2015waa}.

Since the pioneering work by Wigner, Dyson, and Mehta, random matrix theory (RMT) has emerged as a powerful tool to model chaotic dynamics across various complex physical systems, from nuclear spectra to many-body systems \cite{mehta1991random, haake1991quantum, PhysRevLett.52.1}. In the context of quantum chaos, RMT has provided deep insights into the statistical properties of energy levels and their correlation, underpinning phenomena such as spectral rigidity and the universal signatures of chaos in closed and open quantum systems. Notably, random matrix ensembles serve as a bridge to understanding how chaotic dynamics give rise to thermalization—an essential principle encapsulated by the Eigenstate Thermalization Hypothesis (ETH) \cite{Srednicki:1994mfb, Deutch_ETH, DAlessio:2015qtq}. In particular, it becomes crucial to extend RMT beyond traditional ensembles, accommodating systems that exhibit ergodic as well as non-ergodic phases. Such extensions help us to study a versatile range of physical systems, including systems that defy complete thermalization, such as many-body localized states or systems exhibiting slow relaxation dynamics \cite{Pal_MBL, Nandkishore:2014kca}.

One model that exemplifies this broadened perspective is the Rosenzweig-Porter (RP) model \cite{RPmodel}, a prototypical random matrix ensemble with a rich phase structure. The Hamiltonian for the canonical version of this model consists of two terms: a diagonal matrix (whose elements are typically sampled from independent Gaussian distributions) is added to a random matrix drawn from one of the classical Gaussian ensembles. The resulting model is characterized by a single parameter that governs the localization properties of system eigenstates and exhibits three different phases, namely, the ergodic, localized, and fractal phases. In the ergodic phase, eigenstates are extended and uniformly spread across the entire system. Conversely, in the localized phase, eigenstates are confined to a small region, indicating minimal overlap with other states. Additionally, the RP model demonstrates fractal behavior, where eigenstates are neither fully extended nor completely localized, but instead occupy a subset of the available space in a non-uniform manner.

Ever since its proposal as a toy model to explain the behavior of energy levels in atomic spectra \cite{RPmodel}, the RP model has garnered significant interest as an exemplary framework for studying many-body localization, ergodicity, and quantum chaos \cite{BREZIN1996697, Altland_RP, BogomolnyRP, kravtsov2015random}. Experimental findings \cite{Zhang:2023rpm} have demonstrated its suitability in describing spectral transitions from uncorrelated to correlated spectra \cite{Guhr_transition, Guhr2, KunzRP}. Various adaptations of the RP model have been explored in the literature, including different Dyson symmetry classes \cite{Cadez:2024mjc}, generalized versions including circular, log-normal, and L\'evy ensembles \cite{grp1, PoisRP, Buijsman:2021xbi, lognormalRP0, lognormalRP, Monthus_2017, LevyRP, Safonova:2024ffl, LevyRP2, kutlinLevy}, and non-Hermitian generalizations \cite{nonHRP} exhibiting various localization phenomena \cite{madhumitaRP1, madhumitaRP2}. The non-Hermitian case is particularly illuminating, as the presence or absence of the fractal phase depends on whether the diagonal potential is real or complex. Several tools have been employed to investigate this model, such as perturbation theory \cite{Altland_RP}, supersymmetry techniques \cite{Guhr_transition, Guhr2, Truong:2016wcq}, entropic quantities \cite{Pino_2019, lognormalRP0}, adiabatic gauge potentials \cite{Cadez:2024mjc}, Krylov space methods \cite{Bhattacharjee:2024yxj}, the replica approach \cite{Venturelli:2022hka}, and the geometry of Hilbert space \cite{Sharipov:2024lah}.

In recent years, the theory of free probability has emerged as a powerful framework for re-examining quantum chaos from a novel perspective. Originally developed by Voiculescu \cite{voiculescu1986addition, Voiculescu1991, Voiculescu1992} in the context of group isomorphisms and operator algebras, free probability theory was later expanded through its close connections with RMT \cite{speicher2019lecturenotesfreeprobability, mingo, nica2006lectures, Rao_Edelman_2006, edelmanfree}, and began gaining traction in physics literature during the 1990s. Early interest was sparked by the works, among others,  \cite{Gopakumar:1994iq, Douglas:1994zu, Zee:1996qu}, who explored its relevance to large 
$N$ gauge theories and matrix models. More recently, the connection between free probability and generalized versions of the ETH has renewed interest in its role in understanding thermalization in quantum systems \cite{Pappalardi:2022aaz, Pappalardi:2023nsj, Fritzsch:2024qjn, Wang:2025mzl, Vallini:2024bwp, Fava:2023pac}. This resurgence has led to a range of developments of free probabilistic techniques in quantum gravity, such as computing the eigenspectra \cite{Gao:2024lem} of Sachdev-Ye-Kitaev models, or the entanglement spectra in extensions of the ``island'' paradigm \cite{Penington:2019kki, Wang:2022ots}. Notably, recent works \cite{Chen:2024zfj, Camargo:2025zxr} have identified a link between operator statistics and the emergence of asymptotic freeness, while other studies \cite{Jindal:2024zcg} have suggested freeness as a potential missing link between different avatars of quantum chaos.

In this work, we employ an approach based on free probability theory to study the density of states (DOS) and the operator statistics in the RP model. As the system makes a transition from the ergodic phase to the localized phase, the DOS exhibits a transition from a semicircular distribution to a normal distribution. While evaluating the DOS across the three phases of the RP model presents considerable analytical challenges, we develop a semi-analytic approach leveraging free probability tools, including subordination formulae for the resolvent operator. By employing a perturbative method based on the subordination formula, and using the (inverse of the) system size as the perturbation parameter, we derive an approximate DOS in specific regimes of the RP model.  

Although similar techniques have been previously applied to analyze the DOS in various models under constrained settings, our method has the practical advantage that it directly uses the free probability results (which otherwise one usually needs to establish first, \emph{e.g.}, using the replica technique and so on) and is also valid for a general class of Hamiltonians where one of the two terms of the Hamiltonian is Gaussian random matrix. As one of the main results of the paper, we obtain the first two orders of corrections to the Gaussian distribution of the DOS of the RP model when the system is in the fractal phase. By numerical analysis, we show that our approximate analytical expression for the DOS is in very good agreement with it.  

We further study the statistics of the sum of two deterministic spin operators, one of which is time-evolved by the RP model Hamiltonian. At the initial stage, the eigenvalue distribution of the total operator adheres to the \emph{free binomial distribution}. However, as time progresses, the eigenvalue distribution exhibits significant deviations from this initial distribution. Specifically, in the ergodic regime, the eigenvalue statistics of the operator sum show a transition from the \emph{multinomial distribution} to the one predicted by the \emph{free probability distribution} when the component operators are free. The multinomial distribution (discussed later) is the general case of the \emph{binomial distribution}, which holds only for spin-half operators. Conversely, in the fractal and localized regimes, the eigenvalue distributions diverge substantially from the predictions obtained from free probability theory. This observation lends credence to the broader assertion that chaos leads to the asymptotic freeness between two operators at late times. The behavior of operator statistics is consistent with the vanishing contribution of the cumulative higher-point OTOCs at late times.

In the ergodic phase, we quantify the evolution of these distributions over time by systematically comparing them at late times, with respect to key governing parameters. To achieve this, we employ three distinct statistical measures: the $\chi^2$ function, Kullback--Leibler (KL) divergence-- widely recognized as a relative entropy in the context of quantum density matrices-- and the Kolmogorov--Smirnov (KS) test. While the $\chi^2$ function and the divergence rely on computing a notion of statistical distance between two distributions, the KS test is based on the null hypothesis testing. These characterizations are fundamentally different from each other and provide complementary insights, allowing us to define the notion of \emph{free time}: the onset of a timescale where the free probability description arises. Our analysis reveals a clear trend: the \emph{free time} tends to grow progressively as the system transitions from the ergodic phase to the fractal phase. Despite the inherent differences between these statistical measures, they exhibit consistent agreement, underscoring the robustness of our findings.

The present manuscript is structured as follows. Sec. \ref{sec:freeprobability} begins with a brief description of free probability, and we also summarize some known results related to the eigenvalue statistics of the sum of two free operators. We introduce the RP model and its spectral statistics in Sec.\,\ref{RPmodelsection}. It is followed by Sec.\,\ref{DOSfreeprobsection}, where the derivation of the  DOS of the RP model using the tools of free probability is presented. In Sec. \ref{correlationsec}  we compute various $2n$-point correlation functions, which is followed by an in-depth analysis of operator statistics in Sec.\,\ref{sec:operator_stat}. The notion of free time is introduced, and its various properties are quantified. Sec. \ref{concsec} concludes the paper with a summary and some open-ended questions. In four appendices, we elaborate on some further important points. Specifically, in the Appendix \ref{app:ExpGauss} we compare the asymptotic freeness governed by different unitary random matrix ensembles, while in Appendix \ref{app:DecorrEns} we illustrate the role played by the delocalized eigenvectors by considering a decorrelated version of the RP random matrix ensemble. Finally, in Appendix \ref{app:SumSpinS} we review the procedure of obtaining the free convolution prediction for the sum of two spin $s$ operators using the subordination relations, and in Appendix \ref{app:op_dependence} the operator-dependent nature of the emergence free probability prediction is examined.

\section{Elements of Free Probability theory} \label{sec:freeprobability}
Free probability theory is the branch of mathematics that studies non-commutative probability spaces. In this framework, the classical notion of independence between random variables is replaced by the notion of {\it free independence} or {\it freeness} \cite{nica2006lectures, speicher2019lecturenotesfreeprobability, mingo, voiculescu1986addition}. In this section, we provide a brief review of free probability theory and how to use it to characterize chaotic dynamics. 

\paragraph{Non-commutative probability space:}  
A \textit{non-commutative probability space} is a pair $(\mathcal{A}, \varphi)$, where $\mathcal{A}$ is a unital $*$-algebra\footnote{A \textit{unital} algebra $\mathcal{A}$ has a multiplicative identity element $\mathbb{I}$ such that $\mathbb{I}a = a\mathbb{I} = a$ for all $a \in \mathcal{A}$. A \textit{$*$-algebra} is an algebra equipped with an involution $a \mapsto a^*$ satisfying: $(a^*)^* = a$ (involution), $(a + b)^* = a^* + b^*$ (additivity), $(\lambda a)^* = \overline{\lambda} a^*$ for $\lambda \in \mathbb{C}$ (conjugate-linearity), and $(ab)^* = b^* a^*$ (reverses order of multiplication).}
 and $\varphi: \mathcal{A} \rightarrow \mathbb{C}$ is a linear functional satisfying the following properties:
\begin{itemize}
  \item \textbf{$*$-linearity:} $\varphi(a^*) = \overline{\varphi(a)}$\,;
  \item \textbf{Unitality:} $\varphi(\mathbb{I}) = 1$\,;
  \item \textbf{Positivity:} $\varphi(aa^*) \geq 0$ for all $a \in \mathcal{A}$\,;
  \item \textbf{Faithfulness:} $\varphi(a) = 0 \Leftrightarrow a = 0$.
\end{itemize}

\noindent\textbf{Examples}
\begin{enumerate}
  \item \textbf{Quantum mechanics:} A canonical example arises in quantum theory, where $\mathcal{A}$ is the algebra of bounded operators on a Hilbert space $\mathcal{H}$, and $\varphi$ is defined via a quantum state $|\psi\rangle \in \mathcal{H}$ by
  \[
    \varphi(a) = \langle \psi | a | \psi \rangle, \quad \text{for } a \in \mathcal{A}\,.
  \]
  This defines a positive, unital, and $*$-linear functional on $\mathcal{A}$.

  \item \textbf{Random matrix theory:} In this setting, $\mathcal{A}$ consists of random $N \times N$ matrices drawn from an ensemble (\emph{e.g.}, Gaussian Ensembles), and the state is given by the normalized ensemble-averaged trace:
  \[
    \varphi(a) = \frac{1}{N} \mathbb{E}[\operatorname{Tr}(a)]\,.
  \]
  This defines a tracial, positive, and normalized linear functional, which is well-defined in the large-$N$ limit for many standard ensembles of random matrices.
\end{enumerate}

\paragraph{Freeness:} Given a non-commutative probability space $(\mathcal{A}, \varphi)$, we say that two elements $a, b \in \mathcal{A}$ are \textit{free} (or freely independent) if the mixed moments of centered polynomials in $a$ and $b$ vanish whenever the variables alternate and each term is individually centered. More precisely, for all $n \geq 1$ and for any polynomials $\mathcal{P}_1, \mathcal{Q}_1, \dots, \mathcal{P}_n, \mathcal{Q}_n$ in one variable, the following condition holds when $a$ and $b$ are free ~\cite{mingo, speicher2019lecturenotesfreeprobability, nica2006lectures}
\begin{equation} \label{eq-DefFreeness}
\left\langle (\mathcal{P}_1(a) - \langle \mathcal{P}_1(a) \rangle)(\mathcal{Q}_1(b) - \langle \mathcal{Q}_1(b) \rangle) \cdots (\mathcal{P}_n(a) - \langle \mathcal{P}_n(a) \rangle)(\mathcal{Q}_n(b) - \langle \mathcal{Q}_n(b) \rangle) \right\rangle = 0\,.
\end{equation}
where we denote the action of the map by $\varphi(\cdot) = \langle \cdot \rangle$.

In the simplest case where all polynomials are linear, \emph{i.e.}, $\mathcal{P}(x) = \mathcal{Q}(x) = x$, this definition yields:
\begin{align} \label{eq-Producs}
&\langle (a - \langle a \rangle)(b - \langle b \rangle) \rangle = 0\,, \nonumber \\
&\langle (a - \langle a \rangle)(b - \langle b \rangle)(a - \langle a \rangle) \rangle = 0\,, \nonumber \\
&\langle (a - \langle a \rangle)(b - \langle b \rangle)(a - \langle a \rangle)(b - \langle b \rangle) \rangle = 0\,,
\end{align}
and similar relations hold for higher-order terms.
These relations impose specific constraints on joint moments. For example, one finds that:
\begin{equation} \label{eq-Factorization}
\langle ab \rangle = \langle a \rangle \langle b \rangle\,, \quad
\langle aba \rangle = \langle a^2 \rangle \langle b \rangle\,, \quad
\langle abab \rangle = \langle a^2 \rangle \langle b \rangle^2 + \langle a \rangle^2 \langle b^2 \rangle - \langle a \rangle^2 \langle b \rangle^2\,.
\end{equation}
Thus, freeness serves as a non-commutative analogue of statistical independence, ensuring that mixed moments factor in a structured and predictive way, based solely on the individual moments of $a$ and $b$.

\paragraph{Freeness and mixed cumulants:} The implicit condition for freeness in \eqref{eq-DefFreeness} can be expressed more conveniently in terms of special combinations of the moments, known as the free cumulants. The $n$-th (with $n \geq 1$) free cumulant of a random variable $A$  is defined through the following moment-cumulant relation
 \begin{equation}\label{tau_to_kappa}
 	\varphi(A^n) =  \sum_{ \pi \in NC(n)} \kappa_\pi(A), ~~\text{where}~~~~\kappa_\pi (A)= \prod_{b\in \pi} \kappa_{|b|}(A)~,
\end{equation}
where the sum is over all the non-crossing partitions $\pi$ of the set $\{1,\cdots n\}$, denoted as $NC(n)$, and $b$ denotes the blocks of a certain partition $\pi$. Note that the constraint that the sum is over only non-crossing partitions of the set $S=\{1,\cdots n\}$ is the difference between the formula in \eqref{tau_to_kappa} and an analogous formula relating moments and cumulants in the classical probability theory. The total number of partitions  (both crossing and non-crossing ones) of the set $S$ is given by the Bell numbers, the total number of non-crossing partitions is given by the Catalan numbers defined through the relation $C_n= \frac{1}{n+1} \binom{2n}{n}$.

For more than one random variable,  the relation in Eq.\,\eqref{tau_to_kappa} can be generalized to the following
\begin{align}
\begin{split}
	\varphi(A_1 A_2 \cdots A_n) &=  \sum_{ \pi \in NC(n)} \kappa_\pi(A_1, A_2, \cdots A_n)~,\\ \text{where}~~\kappa_\pi (A_1, A_2, \cdots A_n)& = \prod_{b\in \pi} \kappa_{|b|}\Big(
	A_{b(1)}, A_{b(2)}, \cdots A_{b(n)}\Big)~.
\end{split}
\end{align}
Once again, $b=(b(1), b(2), \cdots b(n))$ denotes the blocks of a certain partition $\pi$, and $|b|$ indicates its length. 

With this definition of the free cumulants, one can express the 
condition of freeness between two variables $P$ and $Q$ as the statement that all the mixed free cumulants of these two variables are zero \cite{mingo, nica2006lectures}, \emph{i.e.}, $\kappa_n(C_1, \cdots C_n)=0$ for $n\geq 2$, $C_i \in \{P,Q\}$ for all $i$, and $i$ and $j$ such that $C_i=P, C_j=Q$.  One straightforward conclusion of this fact, along with the relation between the moments and free cumulants, is that all non-crossing moments containing free variables get factorized in terms of moments of the individual variables.

\paragraph{Free Additive Convolution:}

Having discussed the condition that two mutually free random variables, say $a$ and $b$, should satisfy,\footnote{Throughout this paper, we assume that these two variables are self-adjoint and have well-defined eigenvalue density.} we now describe the procedure for obtaining the free probability prediction for the density of eigenvalues of the sum ($a+b$) of these two operators, which will be used in Sec.\,\ref{sec:operator_stat} to compare with numerically obtained distributions. The eigenvalue distribution of $a+b$ is called the free convolution of the eigenvalue distributions $\rho_a(\lambda)$ and $\rho_b(\lambda)$, and is usually denoted as
$\rho_{a+b}(\lambda)=\rho_a(\lambda) \boxplus \rho_b(\lambda)$ \cite{nica2006lectures}. 

The first ingredient we need to define is the resolvent operator associated with the operators $a$ and $b$, which is the Cauchy transformation of the DOS of the respective operator, \emph{e.g.},  
\begin{equation}\label{Cauchy_a}
	G_{a}(z) = \int \frac{\rho_{a}(\lambda)}{z-\lambda} ~ \text{d}\lambda\,.
\end{equation}
Here, the integral is over the support of the distribution $\rho_a(\lambda)$ on the real line. One can recover the distribution $\rho_a(\lambda)$ from the knowledge of the Cauchy transformation $G_{a}(z)$ by using the so-called Stieltjes inversion formula,
\begin{equation}
    \rho_a(\lambda) = \frac{1}{\pi} \lim_{\epsilon \to 0} \left(\Im G_a(\lambda - i\epsilon)\right)\,.
    \label{eq: Stieltjes inversion}
\end{equation}
Next, we notice that the functional inverse of the Cauchy transformation (which we denote here by $\mathcal{B}_a(z)$) is singular at $z=0$, due to the presence of a pole. Therefore, one can obtain its regular part as 
\begin{equation}\label{B_and_R}
    G^{-1}(z) \equiv \mathcal{B}(z)~,~~~  R(z) = \mathcal{B}(z)-\frac{1}{z}\,.
\end{equation}
The quantity $R(z)$ is known as the $R$-transformation of the eigenvalue density.  Before proceeding further, we note two points about the $R$-transformation. Firstly, the domain of dependence of $R_a(z)$ depends on $\rho_a(\lambda)$, and it is well-defined only in a region very close to the origin. Secondly, since the functional inverse $\mathcal{B}(z)$ of the  Cauchy transformation can have multiple branches, to obtain a well-defined  (physical) eigenvalue density, we impose the condition that $R(0)=0$. 

With the above-defined $R$-transformation, to obtain the eigenvalue density of the sum of two operators, we utilize the property that the $R$-transformation of the sum $a+b$ of two self-adjoint operators having compactly supported eigenvalue density is the sum of the $R$-transformations of the individual densities \cite{voiculescu1986addition,nica2006lectures}.\footnote{For derivation of the additivity property of the $R$-transformation for two random matrices, we refer to \cite{Zee:1996qu, zinn1999adding}.} Thus,
\begin{equation}\label{R_addition}
    R_{a+b}(z)=R_a(z)+R_b(z)\,.
\end{equation}
Knowing $R_{a+b}(z)$ from $R_a(z)$ and $R_b(z)$, one can now follow the backward steps to obtain the density $\rho_{a+b}(\lambda)$; namely, first obtain $\mathcal{B}_{a+b}(z)$ from $R_{a+b}(z)$, subsequently invert it to get the corresponding $G_{a+b}(z)$, which can be finally used in the inversion formula  \eqref{eq: Stieltjes inversion} to obtain an expression for the free convolved density $\rho_{a+b}(\lambda)$.

A diagrammatic technique was used in \cite{Zee:1996qu} to derive the additivity property \eqref{R_addition} of the $R$-transformation for the addition of two independent large Gaussian 
random matrices. By noting that the Cuachy transformation $G(z)$ of the density of eigenvalues (\emph{i.e.}, the Green function), can be written in terms of the one-particle irreducible self-energy $\Sigma(z)$ as a Dyson--Schwinger equation $G(z)=1/(z-\Sigma(z))$, we see that the $R$-transformation is just the self-energy, 
\begin{equation}
    \mathcal{B}\big(G(z)\big) = \frac{1}{G(z)}+ \Sigma(z)~~\rightarrow
   ~~ \Sigma(z)= R\big(G(z)\big)\,.
\end{equation}

\subsection{Summary of Known Results on Free Additive Convolution}\label{convolutions}
In this subsection, we present some known results for the free additive convolution of free operators. We refer to \cite{speicher2019lecturenotesfreeprobability,Camargo:2025zxr, nica2006lectures} for a detailed derivation of these results.

\paragraph{Sum of spin-1/2 free operators:}
Consider two free spin-1/2 operators $a$ and $b$ with spectral distribution 
\begin{equation}
\rho_a(\lambda)=\rho_b(\lambda)=\frac{1}{2}\left( \delta(\lambda-1)+\delta(\lambda+1)\right)\,.  
\end{equation}
The spectral distribution of their sum is given by the arcsine distribution
\begin{equation}
    \rho_{a+b}(\lambda)= \begin{cases}
      \frac{1}{\pi}\frac{1}{\sqrt{4-\lambda^2}}\,,~~~ -2 \leq \lambda \leq 2\,.\\
      0\,, ~~~~~~~~~~~~\text{otherwise}\,.
    \end{cases}
    \label{arcsinedist} 
\end{equation}
This distribution is expected to arise for operators of the form $A + B(t)$ in chaotic models where the constituent operators have a spectral distribution described by a Bernoulli distribution. This includes a broad class of systems, such as spin chains and SYK-like models \cite{Camargo:2025zxr}.

\begin{figure}
    \hspace*{-0.3 cm}
    \includegraphics[width=1\linewidth]{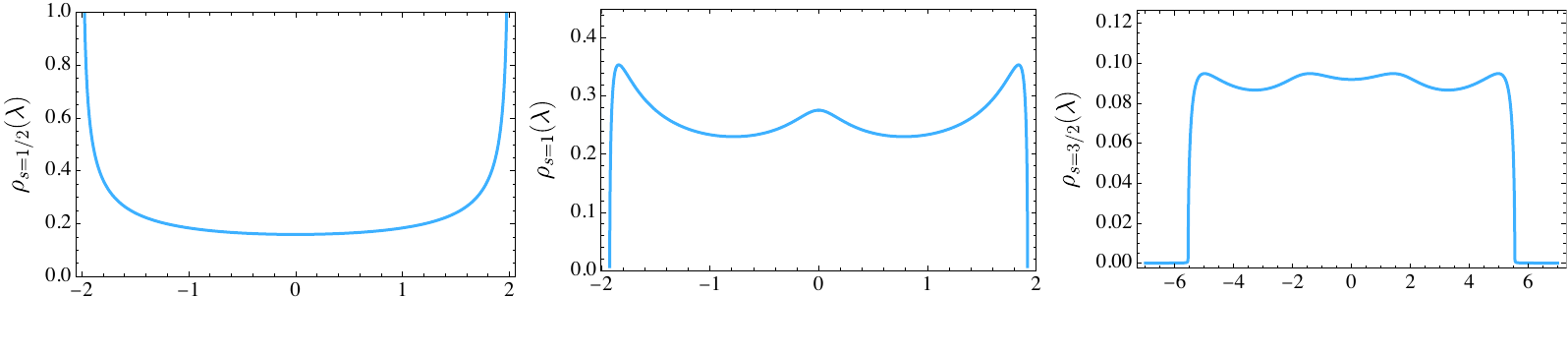}
    \caption{The operator eigenvalue distributions predicted by the free probability for the addition of two free spin $s =1/2$ (left), $s = 1$ (middle), and $s = 3/2$ (right) operators. The spin $1/2$ and the spin $1$ distribution is given by the arcsine distribution \eqref{arcsinedist} and distribution \eqref{eq: sumdis_spin1}, respectively. In contrast, the distribution for spin $3/2$ can be evaluated using the subordination method.}
    \label{fig:allanalplot}
\end{figure}

\paragraph{Sum of spin-1 free operators:}
Consider two free spin-1 operators $a$ and $b$ with spectral distribution 
\begin{equation} \label{eq:multinoulli}
\rho_a(\lambda)=\rho_b(\lambda)=\frac{1}{3}\left( \delta(\lambda-1)+\delta(\lambda)+\delta(\lambda+1)\right) \,. 
\end{equation}
The spectral distribution of their sum is given by the following distribution~\cite{Camargo:2025zxr}
\begin{equation} 
    \rho_{a+b}(\lambda)=\begin{cases}
       \bigg|\frac{ f_1^{2}(\lambda)-4 - 9 \lambda^4 + 33 \lambda^2 }{3 \sqrt{3} \,\pi \lambda \left(\lambda^2-4\right) f_1(\lambda)}\bigg|~,~~~
    -1.9227 \leq \lambda \leq 1.9227\,.\\
0\,,~~~~~~~~~~~~~~~~~~~~~~~~~\text{otherwise}\,.
    \end{cases}
 \label{eq: sumdis_spin1}
\end{equation}
where we have defined the function
\begin{equation}
    f_1(\lambda)=\left( 8+9 \left(3 \lambda^4-30 \lambda^2+70\right) \lambda^2+27 \sqrt{-\lambda^2 \left(\lambda^2-4\right)^2 \left(9 \lambda^4-33 \lambda^2-1\right)}\right)^{1/3}\,.
\end{equation}
The distribution (\ref{eq: sumdis_spin1}) is expected to appear in chaotic models in which the basic operators or composite operators 
have spectral density of the form \eqref{eq:multinoulli}. 

\paragraph{Sum of spin-$s$ free operators:}
For operators with spin larger than unity, it becomes increasingly challenging to obtain analytic formulae for the free additive convolution of the sum of operators, due to the fact that one needs to solve order-$s$ algebraic equations. Fortunately, such distributions can be obtained numerically using subordination function methods; see, for instance~\cite{speicher2019lecturenotesfreeprobability}. Such methods were applied in~\cite{Camargo:2025zxr} for the derivation of the free additive convolution of spin-$3/2$ and spin-$2$ operators. Figure \ref{fig:allanalplot} shows the respective distribution for spin-$1/2$, spin-$1$, and spin-$3/2$. We review the subordination function method in Appendix \ref{app:SumSpinS}.

\subsection{Asymptotic Freeness Produced by Chaotic Dynamics}

Given two operators $P$ and $Q$ that are not initially free, the time evolution of $Q$ under a chaotic Hamiltonian is expected to produce a Heisenberg operator $Q(t) = e^{iHt}\, Q \,e^{-iHt}$ that becomes free from $P$~\cite{Jindal:2024zcg, Chen:2024zfj}. This emergence of freeness defines a novel notion of quantum chaos, which can be incorporated into the quantum ergodic hierarchy~\cite{Camargo:2025zxr}.

Freeness arises only in infinite-dimensional Hilbert spaces. When operators $P$ and $Q$ are represented as $N \times N$ matrices, freeness can emerge only in the limit $N \to \infty$, a regime associated with \textit{asymptotic freeness}. For finite $N$, mixed cumulants never vanish completely and retain residual values at late times. However, under chaotic dynamics -- \emph{i.e.}, when the Hamiltonian exhibits features akin to RMT -- the magnitude of these residual cumulants decreases with increasing $N$, suggesting their vanishing in the $N \to \infty$ limit. In this sense, the finite-$N$ system ``knows'' about its large-$N$ counterpart.

This connection also manifests in operator statistics. Although true freeness between $P$ and the time-evolved operator $Q(t)$ requires infinite $N$, the spectral statistics of $P + Q(t)$ at finite $N$ often agree with free probability predictions. This agreement serves as a practical signature -- or ``smoking gun'' -- of asymptotic freeness.

\subsection{Relation with $2n$-point functions}

It is imperative to ask how the eigenvalue distribution of the sum of two operators is related to the higher-point correlation functions, namely the higher-order OTOCs. To determine the eigenvalue distribution of a matrix $\mathcal{O}$,\footnote{We assume that the operators we consider in this paper have corresponding eigenvalue distributions with compact supports on the real line.} it suffices to evaluate the traces of all powers of the operator, $\text{Tr}(\mathcal{O}^k)$. For a random matrix $\mathcal{O}$, its ensemble-averaged eigenvalue distribution can be derived from $\langle\mathcal{O}^k\rangle$, where $\langle \cdot \rangle$ denotes the infinite-temperature inner product, computed as $\mathbb{E}(\mathrm{Tr}(\cdot))/N$, with $N$ representing the  rank of the matrix \cite{mingo, speicher1997free}. Consequently, the ensemble-averaged eigenvalue distribution of an operator formed as $P + Q$ can be obtained from the moments $\langle(P+Q)^k\rangle$. Therefore, the information about the ensemble-averaged eigenvalue distribution of an operator of the form $P + Q$ can be obtained from the moments $\langle(P+Q)^k\rangle$, or alternatively from the 
mixed moments of the form $\langle P^{n_1}Q^{m_1}P^{n_2}Q^{m_2} \cdots\rangle$.

When one of these operators is a time-evolved Heisenberg operator (as is the case we consider below in Sec. \ref{sec:operator_stat}), the moments  $\langle(P+Q(t)^k\rangle$ have expansions in terms of the OTOCs of different orders.\footnote{We thank Koji Hashimoto for discussions on this point.}  To see this explicitly, consider two spin-$1/2$ operators and the following second-order moment:
\begin{align}
    \langle (P+Q(t))^2 \rangle = \langle P^2 \rangle + \langle Q(t)^2 \rangle + 2 \langle P Q(t) \rangle = 2 + 2 \langle P Q(t) \rangle\,, \label{twopot}
\end{align}
where the normalization $\langle P^{2k} \rangle = \langle Q^{2k} (t) \rangle = 1$ has been applied throughout (keeping in mind they are Pauli operators acting on qubits), and the operators are chosen such that their one-point function vanishes: $\langle P \rangle = \langle Q(t) \rangle = 0$. In other words, we choose the \emph{traceless} operators or the \emph{traceless} part of the operators by appropriately subtracting the traces. For the random matrix ensembles, the average is taken over the ensembles of the Hamiltonians.

Similarly, for the fourth-order moment, we have:
\begin{align}
    \langle (P+Q(t))^4 \rangle = 6 + 2 \langle P Q(t) P Q(t) \rangle\,, \label{fourpt}
\end{align}
where by definition $\langle P^4 \rangle = \langle Q^4 \rangle = 1$. In this expression, the second term can be identified as the lowest-order OTOC. Consequently, Eqs.\,\eqref{twopot} and \eqref{fourpt} reveal that the moments of $P + Q(t)$ encode information of both the two- and four-point OTOCs.

These relationships can be extended to higher-order moments. In chaotic systems, at sufficiently late times, the time evolution is expected to render deterministic operators asymptotically free. Consequently, for operators with vanishing one-point functions, the mixed moments vanish. As a result, the even-order moments simplify under the conditions $\langle P^{2n} \rangle = \langle Q^{2n} \rangle = 1$ and $\langle P^{2n+1} \rangle = \langle Q^{2n+1} \rangle = 0$. Therefore, one obtains
\begin{align}
    \langle (P+Q(t))^{2n} \rangle \simeq \binom{2n}{n}\,, ~~~~~ t \rightarrow \infty\,, 
\end{align}
while all odd moments vanish. 

As expected, these moments correspond directly to those of the free probability prediction for the eigenvalue distribution of the operator $P + Q(t)$ at late times. For example, it is easy to see that the moments of the arcsine distribution in \eqref{arcsinedist} are given by the same binomial coefficients:\footnote{The moment generating function of the arcsine distribution are given by the $I_{0}(2u)$, where $I_n(u)$ is the $n$-th order modified Bessel function of the first kind. It is easy to check that $m_{2n} = d^{2n} I_0(2u)/d u^{2n}\big|_{u \rightarrow 0}$ are the binomial coefficients \eqref{bincoff}, while the odd moments vanish.}
\begin{align}
    m_{2n} = \int_{-2}^2 \mathrm{d} \lambda ~\frac{\lambda^{2n} }{\pi\sqrt{4-\lambda^2}} =  \binom{2n}{n}\,~~~ n = 0,1,2,\cdots\,. \label{bincoff}
\end{align}
Thus, for the sum of two spin-$1/2$ operators, the eigenvalue distribution converges to the arcsine law predicted by the free probability. Extensions to higher spins follow in a similar fashion.

We emphasize that this connection is not a mere coincidence; rather is a direct consequence of the fact that when the time evolution generates free variables, the moments of an operator of the form $P+Q(t)$ should approach at late times those of the distribution predicted by the free additive convolution of the eigenvalue distributions of those of $P$ and $Q$.

\section{Rosenzweig--Porter random matrix ensemble} \label{RPmodelsection}

\subsection{Hamiltonian and the spectral statistics}

The Hamiltonian of the RP model is given by \cite{RPmodel}
\begin{align}\label{HamRP}
    H_{\mathrm{RP}} = A + \frac{1}{N^{\gamma/2}}B\,. 
\end{align}
Here, $A$ is an $N \times N$ random diagonal matrix, chosen from independent and identically distributed (iid) elements from a normal distribution with zero mean and variance $\sigma^2_A = 4/N$. On the other hand, $B$ belongs to the Gaussian Orthogonal Ensemble (GOE) with zero mean and variance $\sigma^2_B = 2/N$. However, the Hamiltonian can be chosen from any three symmetry classes of Dyson ensembles \cite{Cadez:2024mjc}. The Hamiltonian shows rich phase structure across different parametric regimes: the ergodic regime for $0 \leq \gamma \leq 1$, non-ergodic and extended fractal phase for $1 < \gamma < 2$, and localized phase for $\gamma > 2$. The two transitions, namely the ergodic to fractal phase at $\gamma =1$ and the fractal phase to localized phase at $\gamma =2$, can be found analytically at the large $N$ limit. The DOS transitions from a semicircle distribution supported by the interval $[-2, 2]$ for $\gamma = 0$ to the normal distribution with zero mean and variance $4/N$ for large $\gamma$.

The distributions of nearest-level spacings and their ratios constitute interesting diagnostics of chaotic Hamiltonians. Denoting the eigenvalues of the $N \times N$ Hamiltonian as $\lambda_n$ with $n = 1, \cdots, N$, the nearest level spacings are given by the difference between the nearest eigenvalues as $\xi_k = \lambda_{k+1} - \lambda_k$, with $k = 1, \cdots, N-1$. For RMT, such distributions follow universal statistics, given by the Wigner--Dyson distribution \cite{Haake:2010fgh}: 
\begin{align}
    p_{\upbeta}(\xi) =\frac{2 z_{\upbeta} e^{-(\xi z_{\upbeta})^2} (\xi z_{\upbeta})^{\upbeta }}{\Gamma \big(\frac{1+\upbeta}{2}\big)}\,,~~~ 
    z_{\upbeta} =\frac{\Gamma \big(1+ \frac{\upbeta }{2}\big)}{\Gamma \big(\frac{1+\upbeta}{2}\big)}\,,~~~~~ \upbeta = 1, 2, 4\,,
    \label{eq:WD}
\end{align}
where $\upbeta$ is known as Dyson's index. The three standard Dyson ensembles follow the statistics corresponding to $\upbeta = 1$ (GOE), $\upbeta = 2$ (Gaussian unitary ensemble, GUE), and $\upbeta = 4$ (Gaussian symplectic ensemble, GSE), respectively. The Poisson statistics is computed separately and corresponds to the $\upbeta = 0$ limit in the Gaussian $\upbeta$-ensembles \cite{Dumitriu:2002beta, Bujismanbeta}, given by $p_{\upbeta = 0} (\xi) = e^{-\xi}$. These ensembles form a one-parameter family of Gaussian ensembles which reduces to the standard Dyson ensemble for $\upbeta = 1,2,4$.  

However, to compute the level spacing distributions, it is required to locally uniformize the DOS, such that the mean spacing becomes unity $\int_{0}^{\infty} p_{\upbeta}(\xi) \,d \xi = \int_{0}^{\infty} p_{\upbeta}(\xi)\, \xi d \xi = 1$. This method, known as unfolding, often gets complicated, and there is no universal way to perform such a procedure. To avoid this, it is useful to define the distribution of nearest-level spacing ratios, known as the $r$-value statistics, which is free of such unfolding. This is defined as \cite{Oganesyan:2007wpd}:
\begin{align}
    r_n := \frac{\mathrm{min}(\xi_{n+1},\xi_n)}{\mathrm{max}(\xi_{n+1},\xi_n)}\,, ~~~~~~ \langle r \rangle := \mathrm{Mean}(r_n)\,. \label{rratioformula} 
\end{align}
The quantity $\langle r \rangle$ is known as the $r$-ratio, obtained by the mean values of the $r$-value distribution. For Gaussian $\upbeta$-ensembles, the statistical distribution of $r$-values is known \cite{Atas_rvalue}:
\begin{align}
    p_{\upbeta} (r) = \frac{2}{Z_{\upbeta}} \frac{(r + r^2)^{\upbeta}}{(1 + r + r^2)^{1 + (3/2)\upbeta}}\, \Theta(1-r)\,, ~~~~ \upbeta = 1, 2, 4\,, \label{pr}
\end{align}
where $\Theta(1-r)$ is the Heaviside $\Theta$ function restricting the value of $r \in [0,1]$, and the normalization constants are given by $Z_{\upbeta} = 4/27\,~  (\upbeta = 1)$, $2\pi/81\sqrt{3}\,~  (\upbeta = 2)$, and $2\pi/729\sqrt{3}\,~  (\upbeta = 4)$ respectively \cite{Atas_rvalue}. For small $r$, the duistribution $p_{\upbeta}(r) \sim r^{\upbeta}$, which share the similar behavior with the level spacing $(\xi)$ distribution $p_{\upbeta} (\xi) \sim \xi^{\upbeta}$ in the same limit. On the other hand, the large $r$ behavior  $p_{\upbeta}(r) \sim r^{-(2 + \upbeta)}$ is drastically different compared to the rapid exponential decay in $p_{\upbeta} (\xi)$. The Poisson limit is computed separately and given by $p_{\upbeta = 0} (r) = 2/(1+r)^2$.

Besides the distribution itself, it is often convenient to compute its mean $\langle r \rangle_{\upbeta} := \int_{0}^{1} r\, p_{\upbeta} (r) \, dr$. For GOE, GUE, and GSE, it is given by $\langle r \rangle_{\mathrm{GOE}} \approx 0.536$, $\langle r \rangle_{\mathrm{GUE}} \approx 0.603$, and $\langle r \rangle_{\mathrm{GSE}} \approx 0.676$ respectively \cite{Atas_rvalue}. For uncorrelated spectra with Poissonian statistics, it is given by $\langle r \rangle_{\mathrm{P}} \approx 0.386$.

\begin{figure}[t]
\hspace*{-0.2 cm}
\begin{subfigure}[b]{0.5\textwidth}
\centering
\includegraphics[width=\textwidth]{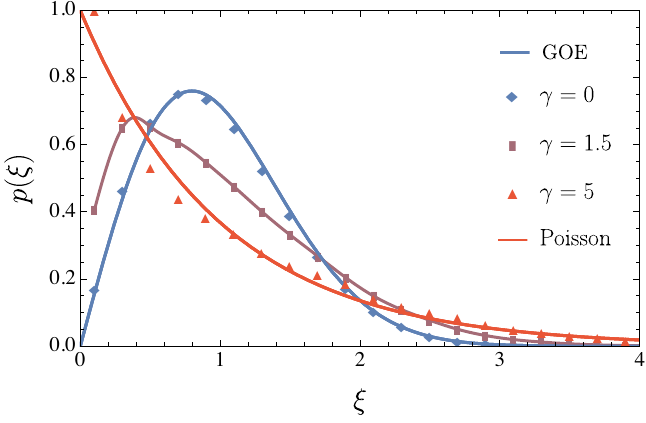}
\caption{Level spacing distribution for different $\gamma$.}
\end{subfigure}
\hfil
\begin{subfigure}[b]{0.5\textwidth}
\centering
\includegraphics[width=\textwidth]{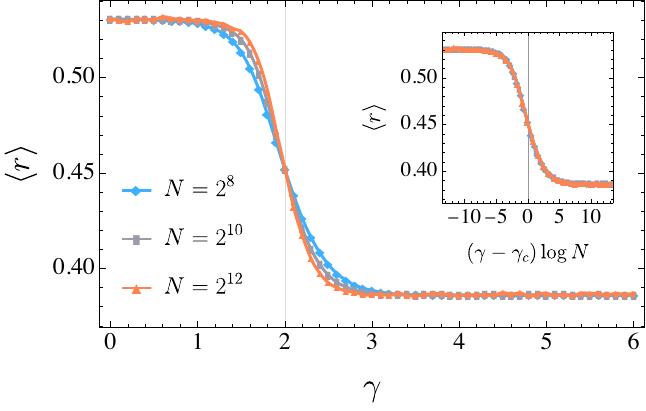}
\caption{Level spacing ratio with $\gamma$.}
\end{subfigure}
\caption{(a) The level spacing distribution $p(\xi)$ in the ergodic ($\gamma = 0)$, fractal ($\gamma = 1.5$), and localized ($\gamma = 5$) regimes. The solid blue and red lines denote the GOE and the Poisson distribution. The system size is $N = 2^{10}$, with a $10^4$ ensemble of Hamiltonians taken. For the spacing, only the bulk of the eigenvalue spectrum ($20\%$ around each side of the mid-spectrum) is considered. (b) The $r$-ratio of the RP model across different phases for different sizes of the matrices $N = 2^8 \,(5 \times 10^4),\, 2^{10} \,(10^4)$ and $2^{12} \,(10^2)$. The ensemble averages are shown in parentheses. The inset shows the data collapse, which clearly marks the fractal to the localized transition at $\gamma_c  = 2$.} \label{fig:eigstatRP}
\end{figure}

Figure \ref{fig:eigstatRP} (a) depicts the level spacing distribution, $p(\xi)$, for the RP Hamiltonian across three distinct phases: ergodic ($\gamma = 0$), fractal ($\gamma = 1.5$), and localized ($\gamma = 5$). The system size is set to $N = 2^{10}$, with $10^4$ independent Hamiltonian realizations. In the ergodic phase, the spacing distribution aligns with the Wigner-Dyson distribution of GOE matrices (Eq.\,\eqref{eq:WD} with $\upbeta = 1$). Conversely, the localized phase exhibits a Poisson distribution, while the fractal phase presents a mixed distribution characteristic.

Additionally, Fig.\,\ref{fig:eigstatRP} (b) illustrates the $r$-ratio for the RP Hamiltonian across various parametric regimes of $\gamma$, encompassing ergodic, fractal, and localized phases. The system sizes chosen are $N = 2^8$, $2^{10}$, and $2^{12}$, with $5 \times 10^4$, $10^4$, and $10^2$ ensembles, respectively. These specific values of $N$ are selected for convenience in later computations involving different spin operators, as described in subsequent sections. The result is consistent with the level spacing distribution in Fig.\,\ref{fig:eigstatRP} (a).

For different system sizes, a clear crossover at $\gamma_c = 2$ is observed, which becomes more prominent after collapsing the data into a single curve. The conclusion drawn is that while the nearest level spacing statistics effectively mark the transition from fractal to localized phases, they fail to adequately capture the transition from ergodic to fractal phases at $\gamma=1$. It is noteworthy that such observations have already been reported in the literature \cite{Pino_2019, PoisRP, Buijsman:2023ips, Bhattacharjee:2024yxj}, and efforts have been made to identify the ergodic to fractal transitions using various probes.

\subsection{Algebra of observables}
In order to fully specify a quantum dynamical system, we need not only the system's Hamiltonian, but also the associated algebra of observables. A generic quantum dynamical system is defined by a pair $(\mathcal{A}, \varphi)$, equipped with a time evolution operator $U(t)$. Given the RP model Hamiltonian $H_\text{RP}$, the time evolution operator is given by $U(t) = e^{-i t H_\text{RP}}$. The Hilbert space of the system is spanned by the eigenvectors of $H_\text{RP}$.

We consider the algebra of observables generated by the spin-$s$ Pauli operators $X_i^{(s)}$, $Y_i^{(s)}$, and $Z_i^{(s)}$, defined in Section \ref{sec:operators}, along with the identity operator on the corresponding Hilbert space.\footnote{Given this initial set of operators, one can construct the associated von Neumann algebra by including limits of sequences of operators that converge in the weak operator topology.} Since we will be interested in the spectral properties of operators and their connection with free probability predictions, we will focus on a map of the form $\varphi(\cdot)=\mathbb{E}\, \text{Tr}(\cdot)/N$, which corresponds to an ensemble average of a infinite temperature thermal state.

\section{Density of states (DOS) from Free Probability} \label{DOSfreeprobsection}

\begin{figure}[h!]
    \centering \includegraphics[width=.85\linewidth]{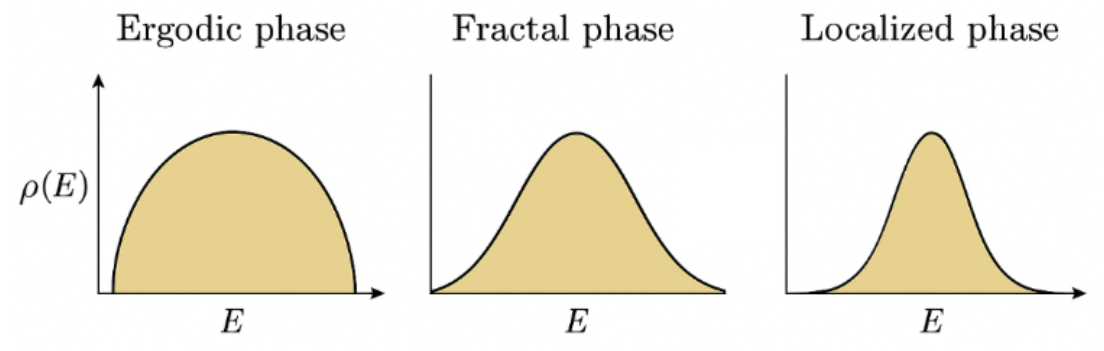}
    \caption{Schematic representation of the DOS in the RP model across its three phases: the ergodic phase (left), characterized by a Wigner semicircle distribution; the fractal phase (center), showing an intermediate form between semicircular and Gaussian; and the localized phase (right), where the DOS is well approximated by a Gaussian distribution.}
    \label{fig:DOSschematic}
\end{figure}

In this section we outline a procedure for approximating the DOS of the RP model \eqref{HamRP} using tools from free probability theory. In the ergodic phase, corresponding to small values of $\gamma$, the DOS is well approximated by the Wigner semicircle distribution. As $\gamma$ increases and the system enters the fractal phase, the DOS gradually deforms, interpolating between the semicircular shape and a Gaussian profile. In the localized phase, which emerges at large $\gamma$, the DOS is well described by a Gaussian distribution. See Fig.~\ref{fig:DOSschematic} for a schematic illustration of this behavior across the three phases. Here, we are interested in analytically computing the DOS specifically for the fractal phase of the RP model.\footnote{For other approaches to the calculation of the DOS of the RP model, we refer to \cite{kreynin1995density, Venturelli:2022hka}.} 

\subsection{Cauchy and R-transforms}

We recall that the functional inverse of the trace of the resolvent (\emph{i.e.}, the Cauchy transformation of the density of eigenvalues), and the $R$-transformation of an operator $\mathcal{O}$ is given by \cite{speicher2019lecturenotesfreeprobability, nica2006lectures}
\begin{equation}\label{B_and_R1}
	G_{\mathcal{O}}^{-1}(z) \equiv \mathcal{B}_{\mathcal{O}}(z) = \frac{1}{z} + R_{\mathcal{O}}(z)\,, 
\end{equation}
with the expression for the resolvent in terms of its DOS $\rho_{\mathcal{O}} (\lambda)$ as 
\begin{equation}\label{Cauchy_A}
	G_{\mathcal{O}}(z) = \int \frac{\rho_{\mathcal{O}}(\lambda)}{z-\lambda} ~ \text{d}\lambda~.
\end{equation}
%Here, the integral is over the support of the distribution $\rho_{\mathcal{O}} (\lambda)$. Note that $G_{\mathcal{O}}(z) $ is essentially the Cauchy transformation of $\rho_{\mathcal{O}} (\lambda)$. 

When the operator $\mathcal{O}$ is the Hamiltonian of a quantum mechanical system belonging to a random matrix ensemble, one can relate the resolvent to the time evolution operator as follows. Consider the trace of the ensemble-averaged time evolution operator:
\begin{equation}
    \mathcal{U}_H(t) = \frac{1}{N}\mathbb{E}\big(\text{Tr} ~e^{-i H t}\big)\,,
\end{equation}
which is related to the DOS through a Fourier transformation \cite{BREZIN1996697}
\begin{equation}\label{dos_FT}
    \rho_{H}(\lambda)= \frac{1}{2 \pi}\int_{-\infty}^{\infty} \text{d}t~e^{-i t \lambda}~\mathcal{U}_H(t)\,.
\end{equation}
From the definition of the resolvent in \eqref{Cauchy_A}, it is easy to see that $G_H(z)$ is related to the averaged time evolution operator as \cite{BREZIN1996697}
\begin{equation}\label{G_evolution}
    G_{H}(z) = i  \int_{0}^{\infty} ~ \text{d}t ~ \mathcal{U}_H(t) ~e^{-itz}\,.
\end{equation}

\subsection{Addition of two free operators: subordination formulae for the resolvent}

Let us consider two \emph{free} operators $P$ and $Q$ which have well-defined DOS in the limit $N \rightarrow \infty$, and we are interested in the DOS of the sum of the operators $C=P+Q$. From the additivity of the $R$ transformation: $R_C = R_P + R_Q$ \cite{speicher2019lecturenotesfreeprobability, nica2006lectures, voiculescu1986addition}, we write
\begin{equation}
	\mathcal{B}_C(z) = R_{C}(z)+\frac{1}{z}\,,~~ \Rightarrow ~~ \mathcal{B}_P(z) = \mathcal{B}_C(z)- R_{Q}(z)\,.
\end{equation}
Substituting $z \rightarrow G_{C}(z)$ in the argument of the last expression above, we find
\begin{equation}
	\mathcal{B}_P\big(G_{C}(z)\big) = z- R_{Q} \big(G_{C}(z)\big)\,,
\end{equation}
where we have used $\mathcal{B}_C(G_{C}(z)) = z$, since $G_C$ and $\mathcal{B}_C$ are functional inverses to each other. Further applying $G_{P}$ on both sides and using the same functional inverse relationship, we get a self-consistency equation for $G_{C}(z)$:
\begin{equation}\label{subor_A}
	G_{C}(z) = G_{P}\big(z-R_{Q}\big(G_{C}(z)\big) \big)\,.
\end{equation}
One can perform an entirely analogous procedure for the operator $Q$ to obtain
\begin{equation}\label{subor_B}
	G_{C}(z) = G_{Q}\big(z-R_{P}\big(G_{C}(z)\big) \big)~.
\end{equation}
In the following, we use these formulae to obtain an expression for the DOS for the RP model. The selection between the above two formulae \eqref{subor_A} and \eqref{subor_B} for determining the eigenvalue density of $C$--which can be extracted from $G_{C}(z)$ through the Stieltjes inversion formula--is guided by the relative simplicity of applying the resolvent and the $R$-transformation to either operator $P$ or $Q$.

Before proceeding further, we observe that a self-consistency equation, analogous to the one in \eqref{subor_A} for the resolvent of the RP model, was derived in \cite{Venturelli:2022hka} through the replica trick. The authors also discussed the relationship between their formula and that obtained via the law of addition in free probability. The following discussion, which addresses the approximate evaluation of the DOS for the RP model, closely parallels the treatment in Ref.\,\cite{Venturelli:2022hka}. Nevertheless, the exposition has been kept sufficiently general to encompass not only the RP model but also the addition of two non-commutative operators, one of which exhibits a Wigner semicircle as its DOS.

\subsection{A perturbative scheme for the resolvent}
\label{sec:PertSchemeResolv}

For most choices of the operators $P$ and $Q$, the expressions for their respective resolvents are sufficiently intricate, making it challenging to solve either of the equations in Eqs.\,\eqref{subor_A} or \eqref{subor_B} for $G_C(z)$. Consequently, we proceed to derive a perturbative expression for the DOS of an operator of the form $C = P + \alpha Q$, where $P$ and $Q$ are mutually free, and their respective DOS expressions are assumed to be known. Here, $\alpha$  is a parameter which we assumed to be small, thereby allowing us to develop a perturbation series for the Cauchy transformation of the DOS of the operator $C$. For the RP model Hamiltonian, we have $ \alpha = N^{-\gamma/2}$.  

Utilizing the scaling transformation rule of the $R$-transformation,\footnote{The $R$-transformation of the density of eigenvalues for an operator $P$ obeys the scaling law: $R_{\alpha P}(z) = \alpha R_P(\alpha z)$, where $\alpha$ is a constant.} along with the definition given in Eq.\,\eqref{Cauchy_A}, we obtain, from Eq.\,\eqref{subor_A}: 
\begin{align}\label{GC_exact}
	G_{C}(z) &=  \int \text{d}\lambda~ \frac{\rho_{P}(\lambda)}{z-\alpha R_{Q} \big(\alpha G_{C}(z)\big)-\lambda} \nonumber \\
    &=i  \int_{0}^{\infty} ~ \text{d}u ~\chi_P(-u)  \exp \Big[
	-i u \Big(z- \alpha R_{Q} \big(\alpha G_{C}(z)\big) \Big)\Big]~.
\end{align}
Here we have defined the characteristic function  $\chi_P(u)$ of the distribution $\rho_{P}(\lambda)$ by a Fourier transformation 
\begin{equation}
\chi_P(u) := \int \text{d}\lambda ~\rho_{P}(\lambda)~ e^{- i \lambda u }\,. \label{chiu}
\end{equation}
Comparing this expression with the inverse of Eq.\,\eqref{dos_FT}, we see that the characteristic function is nothing but the ensemble-averaged 
evolution operator generated by $P$, \emph{i.e.}, $\chi_P(-t)=\mathcal{U}_{P}(t)$. 

Next, expanding the exponential factor containing $R_{Q}$ in Eq.\,\eqref{GC_exact} we have the following series expansion for $G_{C}(z)$:
\begin{equation}\label{GC_RB}
	G_{C}(z) = i  \int_{0}^{\infty} ~ \text{d}u ~\chi_P(-u) ~e^{-iuz}  \sum_n \frac{(iu \alpha)^n}{n!} \big(R_{Q} \big(\alpha G_{C}(z)\big)\big)^n\,.
\end{equation}
Now, we express the $R$-transform of any operator in terms of \emph{free cumulants} $\kappa_j$, defined by the following relation \cite{mingo}:
\begin{align}
    R(z) := \sum_{j=1}^{\infty} \kappa_j z^{j-1}\,. \label{rz}
\end{align}
Hence, expressing in terms of the free cumulants $\kappa_j(Q)$ of the operator $Q$, $G_C(z)$ has the following generic form
\begin{equation}\label{GC_kappa}
	G_{C}(z) = \sum_n \frac{i^{n+1}\alpha^n}{n!} \Big(\sum_{j=1}^{\infty}\kappa_j(Q)\big(\alpha G_{C}(z)\big)^{j-1}\Big)^n  \int_{0}^{\infty} ~ \text{d}u ~\chi_P(-u) ~ u^n ~e^{-iuz}\,.
\end{equation}

So far, the above expression for $G_c(z)$ is exact. For any generic operator $Q$, its $R$-transformation can be an arbitrary function of $z$, so that it is difficult to obtain a perturbation series with $\alpha$ as a small parameter. However, one particular case where the above expression becomes easily tractable is when the operator $Q$ has Wigner semicircle as the DOS (which is the case also for the RP model considered in this paper), \emph{i.e.}, $\rho_{Q}(\lambda) = \frac{1}{2 \pi} \sqrt{4-\lambda^2}$. In that case\footnote{The Wigner semicircle is the analogue of the Gaussian distribution for non-commutative probability. Its $R$-transform takes a particularly simple form: $R(z) = z$. Given that $R(z)$ generates free cumulants via the relation \eqref{rz}, the semicircle distribution is characterized by a single non-vanishing free cumulant, which is the second free cumulant and for the semi-circle distribution above, it has a value of unity.} $R(z)=z$, and the expression in Eq.\,\eqref{GC_RB} (or \eqref{GC_kappa}) simplifies to 
\begin{align}
  G_{C}(z) &=  \sum_n\frac{i^{n+1}}{n!} \alpha^{2n}\,\big(G_{C}(z)\big)^n \int_{0}^{\infty}  \text{d}u ~u^n \chi_P(-u) e^{-iuz} \,,\nonumber \\
  &=\sum_n \frac{(-1)^n}{n!} \alpha^{2n}\, \big(G_{C}(z)\big)^n\, \partial_z^n G_{P}(z)\,,
\end{align}
where to obtain the last expression, we have used an analogous relation to that of the one in \eqref{G_evolution}, valid for the operator $P$, \emph{i.e.}, $G_P(z)= i \int_0^{\infty} \chi_P (-u)\, e^{-i u z}$, with $\chi(u)$ given by Eq.\,\eqref{chiu}.

Evaluating the terms of the above sum recursively, one can obtain the resolvent $G_C(z)$ with corrections of different orders of $\alpha$. Thus, at $\mathcal{O}(\alpha^0)$, we have 
\begin{equation}
	G_{C}^{(0)}(z) = i  \int_{0}^{\infty} ~ \text{d}u ~\chi_P(-u) ~e^{-iuz}~=G_{P}(z)\,.
\end{equation}
The expressions for $G_C(z)$ with first and second order corrections  are, respectively, 
\begin{equation}\label{GC_1}
	G_{C}(z) \approx G_{C}^{(1)}(z) + \mathcal{O}(\alpha^4)~,~~\text{where}~~ ~~G_{C}^{(1)}(z) ~= G_{C}^{(0)}(z) - \alpha^2  G_{C}^{(0)}(z) ~ \partial_z G_{C}^{(0)}(z) ~.
\end{equation}
and $G_{C}(z) \approx G_{C}^{(2)}(z) + \mathcal{O}(\alpha^6)$ where
\begin{align}\label{GC_2}
G_{C}^{(2)}(z) &= G_{C}^{(0)}(z) - \alpha^2  G_{C}^{(0)}(z) \partial_z G_{C}^{(0)}(z)\, \nonumber \\
&+ \frac{\alpha^4}{2} \Big(G_{C}^{(0)}(z) - \alpha^2  G_{C}^{(0)}(z) \partial_z G_{C}^{(0)}(z)\Big)^2 \partial_{zz} G_{C}^{(0)}(z)\,.
\end{align}
Hence, all correction terms can be expressed in terms of $G_{C}^{(0)}(z) = G_P(z)$ and its derivatives with respect to $z$. These, in turn, can be derived from the characteristic function of the distribution $\rho_{P}(\lambda)$. Furthermore, owing to the specific scaling behavior of the $R$-transformation, the $n$-th order correction to $G_{C}^{(0)}(z)$ is proportional to $\alpha^{2n}$. The approximate DOS for the operator $C$ can then be determined from $G_{C}(z)$ using the Stieltjes inversion formula \eqref{eq: Stieltjes inversion}
\begin{equation}\label{dos_from_G}
	\rho_{C}(\lambda) = \frac{1}{\pi}  \lim_{\epsilon \rightarrow 0} (\text{Im} ~ G_{C}(\lambda- i \epsilon))\,.
\end{equation}

\begin{figure}[t]
\centering
\includegraphics[width=0.6\textwidth]{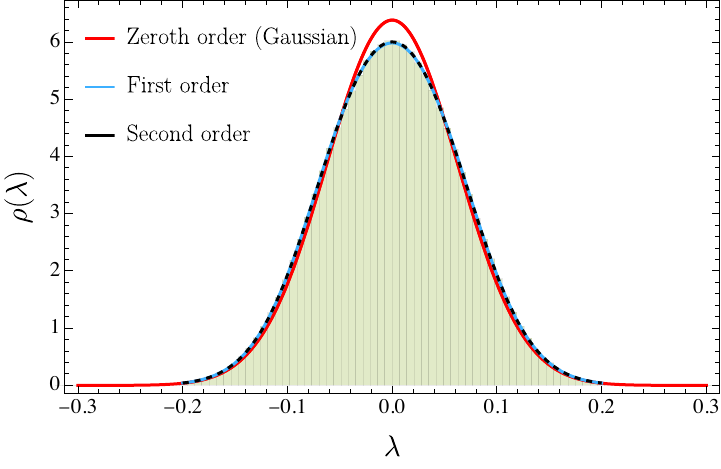}
\caption{DOS of the RP model: comparison of approximate analytical 
expression with numerical results. The blue and black-dashed curves respectively represent the analytical DOS with first and second-order corrections. The red curve represents the Gaussian eigenvalue distribution of the matrix $A$. Here we have set $s=1/2,~\gamma=1.1, L=10$, and the histogram shows the numerical DOS averaged over 5000 independent realisations of the RP model Hamiltonian.} 
\label{fig:DOS_RP_APPROX}
\end{figure}

\subsection{Approximate DOS of the RP model}
\label{sec:ApproxDOSforRP}
Thus far, the derived approximate expressions for $G_C(z)$ hold for any operator $P$ with a well-defined DOS, combined with an operator $Q$ characterized by a semicircular density. Turning now to the specific case of the RP model examined in this paper, the eigenvalue distribution of the diagonal matrix $P \equiv A$ follows a Gaussian distribution with zero mean and variance $\sigma^2 = 4/N$, while $Q \equiv B$. Consequently, we identify the operator $C$ as the Hamiltonian $H$ of the RP model. Using the Gaussian DOS of $P$ as $\rho_P(\lambda) = \frac{1}{\sqrt{2 \pi  \sigma^2}}  e^{-\frac{\lambda^2}{2 \sigma^2}}$, we find $\chi_P(u) = e^{-\frac{1}{2}u^2 \sigma^2}$, leading to the following zeroth-order resolvent of operator $C$: 
\begin{equation}
	G_{C}^{(0)}(z) =  \sqrt{ \frac{\pi}{2 \sigma^2}} \Bigg[i e^{-\frac{z^2}{2 \sigma^2}} +  \frac{2}{\sqrt{\pi}} D \bigg(\frac{z}{\sqrt{2} \sigma}\bigg)\Bigg]~,
\end{equation}
where $D(x)$ denotes the Dawson function: $D(x) = e^{-x^2} \int_{0}^{x}e^{t^2}~\text{d}t$.  Using this expression for $G_{C}^{(0)}(z)$ in Eqs.\,\eqref{GC_1} and \eqref{GC_2}, we obtain the first and second-order corrections to $G_C(z)$, which, in this scenario, can also be expressed in terms of the Dawson function. Consequently, the DOS of the RP model can be obtained from \eqref{dos_from_G}. It is important to note that, for the RP model, this approximation for the DOS holds when $\gamma > 1$.\footnote{Even though our formalism is general, the formula above captures only small corrections to the Gaussian distribution, which is characteristic of the localized phase. Naturally, such corrections might still provide an accurate description of the DOS in the fractal phase--an observation consistent with our findings. For $\gamma \lesssim 1$, however, the DOS resembles a deformed semicircle, and we do not expect the formula to yield quantitatively accurate predictions in this regime.}

In Figure \ref{fig:DOS_RP_APPROX}, we have shown the DOS of the RP model obtained using this procedure along with the one obtained numerically.\footnote{In Ref.\,\cite{Truong:2016wcq}, an implicit formula for the DOS of a Hamiltonian where a random matrix perturbation ($V$) is added to a random (or a non-random matrix) matrix $H_0$ was derived using supersymmetric techniques, including the case of the RP model. The numerical results reported there seem to be consistent with the ones reported here.} As can be seen, the perturbative expression, even at $\mathcal{O}(\alpha^2)$, has an excellent agreement with numerical results.

\section{Behavior of $2n$-point correlators} \label{correlationsec}

\subsection{Choice of operators}

The $2n$-point correlation functions of an operator $\mathcal{O}$ with appropriate normalization that we are interested in are defined as 
\begin{align}
    F_{2n, (s)} (t) =  \langle (\mathcal{O}^{\dagger}(0) \mathcal{O}(t))^n \rangle = \frac{1}{(2s+1)^L} \mathbb{E}\,\mathrm{Tr} \big[(\mathcal{O}^{\dagger}(0) \mathcal{O}(t))^n \big] \,, ~~n = 0, 1, 2, \cdots\,,
\end{align}
where $\mathcal{O}(t) = e^{i H t} \mathcal{O}(0) e^{-i H t}$ is the time-evolved operator and $\langle \cdot \rangle = \mathbb{E}(\mathrm{Tr}(\cdot))/(2s+1)^L$. Here, $\mathbb{E}$ refers to the ensemble average, assuming the Hamiltonian $H$ is drawn from a random matrix ensemble. For the RP model, the averaging encompasses the distributions of both operators $A$ and $B$ in \eqref{HamRP}.

By definition, $F_{0, (s)} (t) = 1$ for all time. For example, the $2$-point and $4$-point functions are given for $n = 1$ and, $n = 2$ respectively
\begin{align}
    F_{2, (s)} (t) &= \frac{1}{(2s+1)^L} \mathbb{E}\,\mathrm{Tr} \big(\mathcal{O}^{\dagger}(0) \mathcal{O}(t) \big) \,,\\
    F_{4, (s)} (t) &= \frac{1}{(2s+1)^L} \mathbb{E}\,\mathrm{Tr} \big(\mathcal{O}^{\dagger}(0) \mathcal{O}(t) \mathcal{O}^{\dagger}(0) \mathcal{O}(t) \big)\,.
\end{align}
The $4$-point functions are commonly known as the out-of-time-ordered correlator (OTOC). Since we will be focusing on the unitary evolution of the Hermitian operators, we have $\mathcal{O}^{\dagger}(t) = \mathcal{O}(t)$ for all time.

We define the operators as tensor products of spin operators with the following explicit forms:
\begin{align}
    S_1 &\equiv S_1(0) =  \epsilon \sqrt{\frac{s(s+1)}{3}} \, Z_1^{(s)} \otimes I_2 \otimes \cdots \otimes I_{L}\,, \label{aop}\\
     S_L &\equiv S_L(0) =  \epsilon \sqrt{\frac{s(s+1)}{3}}\, I_1 \otimes \cdots \otimes I_{L-1} \otimes  Z_L^{(s)}\,, \label{bop}
\end{align}
where the spin system consists of $L$ lattice sites, with $Z_k^{(s)}$ and $I_k$ being Pauli $\sigma_z$ matrix (of spin $s$) and identity operators acting on site $k$. The parameter $\epsilon$ distinguishes between spin types: $\epsilon = 2$ for half-integer spins and $\epsilon = 1$ for integer spins. 

The operators are normalized using the Frobenius inner product: $\langle P|Q \rangle = \mathrm{Tr}(P^{\dagger} Q)/(2s+1)^L$, ensuring the dimensional consistency. Here, $S_k$ refers to spin operators corresponding to spin $s$, with dimension $(2s+1)^L$. The spin operators adhere to the following fundamental relations, which govern their commutation properties and additive structure acting on site \cite{Craps:2019rbj}
\begin{align}
    &[X_j^{(s)}, Y_k^{(s)}] = i \sqrt{\frac{3}{s(s+1)}} \, \delta_{jk} \, Z_j^{(s)}\,,\\
    &X_i^{(s)} X_i^{(s)} + Y_i^{(s)} Y_i^{(s)} + Z_i^{(s)} Z_i^{(s)} = 3\, \mathbb{I}_{N}\,, ~~~~ (i~\mathrm{not~summed}).
\end{align}
where $\mathbb{I}_N$ represents the identity matrix of dimension $N = (2s+1)^L$, same as the dimension of the RP model Hamiltonian. For the special case of $s = 1/2$, the operators reduce to the Kronecker products of Pauli matrices, thereby aligning with the familiar formalism of spin $1/2$.

Using the above definition, we study the behavior of the $2n$-point OTOCs:
\begin{align}\label{eq:npointfunctions}
     F_{2n, (s)}^{(\gamma)} (t) = \frac{1}{(2s+1)^L} \mathbb{E}\, \mathrm{Tr} \big[(S_i(0) S_j(t))^n \big] = \langle (S_i(0) S_j(t))^n \rangle \,, ~~n = 0, 1, 2, \cdots\,,
\end{align}
throughout the regime $\gamma \in [0,1]$, \emph{i.e.}, in the ergodic phase. From now on, we concentrate on operators with spin $s=1/2$ and omit the subscript $s$.

\subsection{Correlation functions in the ergodic phase from asymptotic freeness}
Using the fact that, deep in the ergodic phase of the RP model, the operator dynamics resembles that of Haar-random evolution, we can exploit the notion of freeness between random and deterministic matrices in the large-dimension limit to derive explicit expressions for the $2n$-point correlation functions. First, consider the two-point correlation function $\langle S_i(0)S_i(t) \rangle $, which can be expanded as  
\begin{align}\label{2pt_free1}
    \langle S_i(0)S_i(t) \rangle= \sum_{k,l} \frac{(it)^{k}(-it)^{l}}{k!~ l!} \langle{S_i H^k S_iH^l}\rangle\,.
\end{align}
To evaluate the mixed moment $\langle S_i H^k S_i H^l \rangle$, we use the fact that, deep in the ergodic phase, the Hamiltonian $H$ can be treated as free from the deterministic operator $S_i$ in the limit $N=(2s+1)^L \rightarrow \infty$. Since $S_i$ is traceless, the only non-vanishing contribution in the moment expansion comes from the diagram that factorizes as $\langle{S_i H^k S_iH^l}\rangle~=\langle{S_i  S_i}\rangle \langle H^k \rangle \langle H^l \rangle$ \cite{Camargo:2025zxr}. 

Next, to calculate moments of the Hamiltonian of different orders, we need the DOS of the Hamiltonian. To this end, note that, deep in the ergodic phase of the RP model (with sufficiently small values of the parameter $\gamma$) and for large $N$, the normalized DOS is well approximated by a Wigner semicircle \cite{kravtsov2015random, guhr1996transitions}:
\begin{align} \label{eq:WignerR}
\rho(E) = \frac{2}{\pi R^2} \sqrt{R^2 - E^2}\,,
\end{align}
where $R$ is the radius of the semicircle distribution. For the RP Hamiltonian, in the regime of large $N$ and small $\gamma$, the radius scales approximately as $R = 2/N^{\gamma/2}$, consistent with the structure of the Hamiltonian.\footnote{For finite $N$, the semicircle formula describes the DOS up to some $\gamma \leq \gamma_{*}$ with $\gamma_{*} \geq 0$ in the ergodic phase. For example, with $N=500$, we observed that the Wigner semicircle describes the DOS well as long as $\gamma \leq 0.5$.}

\begin{figure}[t]
\hspace*{-0.3 cm}
\includegraphics[width=1\textwidth]{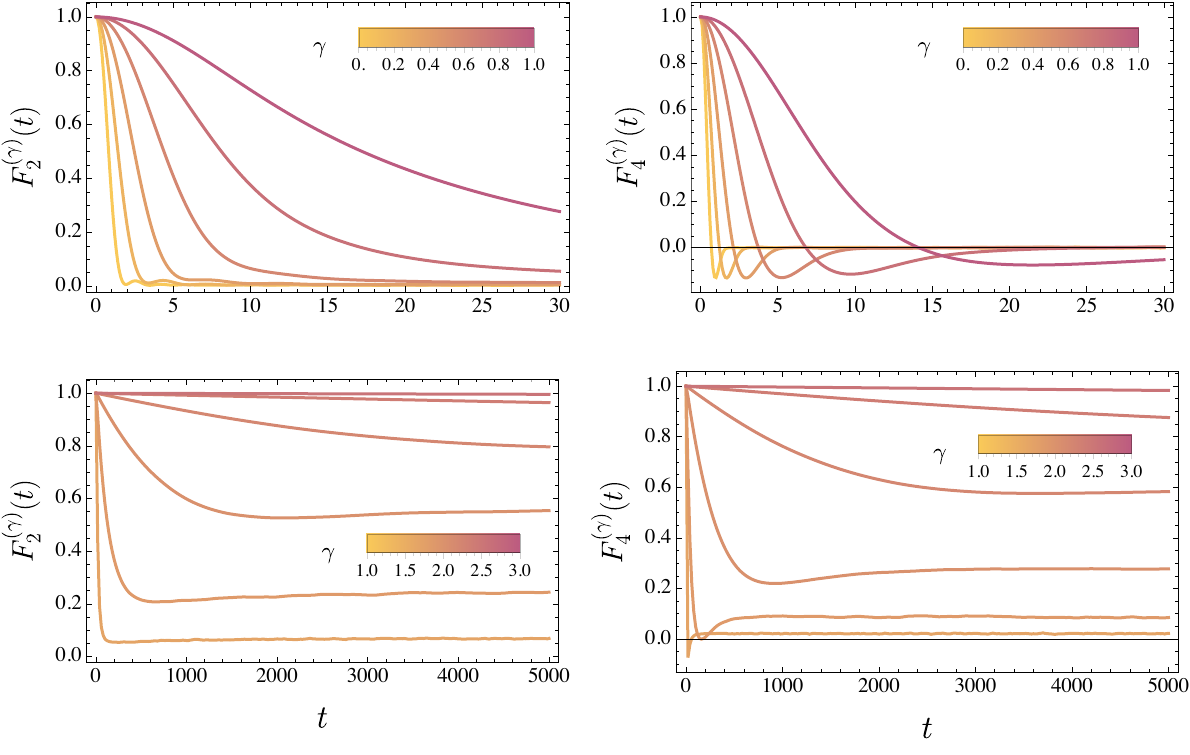}
\caption{Two-point function $F_2^{(\gamma)}(t)$ (left) and four-point OTOC $F_4^{(\gamma)}(t)$ (right), shown in the ergodic phase (top row) and in the non-ergodic phases (bottom row). Parameters are set to $i=j=1$, $L=8$, and results are averaged over 100 realizations of the Hamiltonian.} \label{fig:F2andF4}
\end{figure}

For even values of $n$, the moments of the Hamiltonian obtained from the above DOS are given by $\langle H^{n} \rangle =C_n (R/2)^{2n}$, while for odd $n$, the moments vanish. Here, $C_n$ are Catalan numbers. Using \eqref{2pt_free1}, the two-point function can be expressed as
\begin{equation}\label{2pt_free}
    \langle S_i(0)S_j(t) \rangle= \langle{S_i S_j}\rangle~ \sum_{k,l} \frac{(it)^{2k}(-it)^{2l}}{k!~ l!} C_{2k} C_{2l} \Big(\frac{R}{2}\Big)^{2(k+l)}= \langle{S_i S_j}\rangle \left( \frac{2 J_1(Rt)}{Rt} \right)^2\,.
\end{equation}
Since the ensemble-averaged analytically continued partition function at infinite temperature is
\begin{align}
    \langle Z(t) \rangle= \int \textrm{d}E\, \rho(E)~ e^{-iEt}~=\frac{2 J_1(Rt)}{Rt}\,.
\end{align}
The time-dependent term on the right-hand side of the final expression in \eqref{2pt_free} corresponds to the part of the spectral form factor arising from the disconnected contribution of the two-point correlation function. For $\gamma = 0$, the expression for $\langle Z(t) \rangle$ reduces to the Gaussian (\emph{e.g.}, GOE) result:  $J_1(2t)/t$, as expected. 

Next, we consider the four-point OTOC. For simplicity, we focus on the case of initially disjoint operators with zero one-point functions. This correlation function can be written as 
\begin{equation}\label{4pt_free}
    \langle S_i(0)S_j(t)S_i(0)S_j(t) \rangle= \sum_{k,l,m,n} \frac{(it)^{k+m}(-it)^{l+n}}{k!~ l!~m!~n!} \langle{S_i H^k S_jH^lS_i H^m S_jH^n}\rangle\,.
\end{equation}
Since the initial operators are assumed to be non-overlapping, only a single diagram gives a non-zero contribution in the moment expansion in the large-$N$ limit, due to the freeness between the Hamiltonian $H$ and the operators $S_i$ (and $S_j$) in the ergodic phase of the RP model. Proceeding similarly to the two-point function, this contribution can be evaluated as \cite{Camargo:2025zxr}:
\begin{align}\label{OTOC_free}
  \langle S_i(0)S_j(t)S_i(0)S_j(t) \rangle=  \langle{S_i S_jS_i S_j}\rangle \left( \frac{2 J_1(Rt)}{Rt} \right)^4~,~~i \neq j~.
\end{align}
Once again, the final expression corresponds to the disconnected contribution of the spectral correlation function to the four-point spectral form factor \cite{Cotler:2017jue}. 

\begin{figure}[t]
\centering
\includegraphics[width=.6\textwidth]{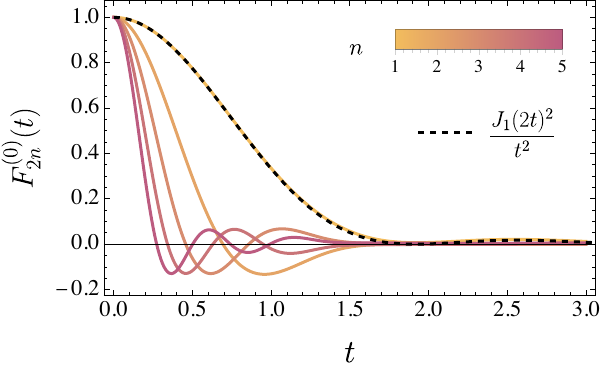}
\caption{Behavior of the $2n$-point function for $n = 1$ to $n = 5$ in the ergodic phase ($\gamma = 0$), governed by a fully GOE Hamiltonian. The black dashed line is the analytic result \eqref{eq:F2_analytic} with $R = 2$ (\emph{i.e.}, $\gamma = 0$). Here, we set $i=j=1$, $L=8$, and average over 100 realizations of the Hamiltonian.} \label{fig:F2n}
\end{figure}

\paragraph{Simplifications for the spin-1/2 case} 
For $s=1/2$, expressions \eqref{2pt_free} and \eqref{4pt_free} can be further simplified using 
$\langle S_i S_j\rangle = \delta_{ij}$ and $\langle S_i S_j S_i S_j\rangle = \langle S_i^2 S_j^2\rangle = 1$ for $i \neq j$, 
yielding the simple results:
\begin{align}
    F_2^{(\gamma)}(t) &= \langle S_i(0) S_j(t) \rangle = \delta_{ij}\,\left( \frac{2 J_1(R t)}{R t} \right)^2 \,,\label{eq:F2_analytic}  \\
     F_4^{(\gamma)}(t) &= \langle S_i(0) S_j(t) S_i(0) S_j(t) \rangle = \left( \frac{2 J_1(R t)}{R t} \right)^4\,, \quad i \neq j \,. \label{eq:F4_analytic}
\end{align}
We remind the reader that the expressions for the correlation functions derived above using free probability theory are valid strictly in the $N \rightarrow \infty$ limit. For finite $N$, numerical results may exhibit deviations from these predictions; however, these differences vanish as $N$ becomes large \cite{Camargo:2025zxr}.
   
Figure \ref{fig:F2andF4} shows the behavior of the two-point function
$F_2^{(\gamma)}(t)$ (Eq.\eqref{eq:npointfunctions} with $n=1$ and $i=j=1$) and the four-point OTOC $F_4^{(\gamma)}(t)$ (Eq.\eqref{eq:npointfunctions} with $n=2$ and $i=j=1$) across the ergodic and non-ergodic phases for $s=1/2$. In the top row, both $F_2^{(\gamma)}(t)$ and $F_4^{(\gamma)}(t)$ decay to zero in the ergodic phase, with the decay becoming progressively slower as $\gamma$ increases. In contrast, the bottom row shows that for $\gamma>1$, the correlation functions no longer decay to zero. These observations support the picture that, in the ergodic phase, the algebra of observables at $t=0$ becomes \emph{asymptotically free} from the algebra of observables at later times, with the apparent timescale for the \emph{onset of freeness} or \emph{free time} (to be precisely defined later) growing with $\gamma$. By contrast, in the non-ergodic phases, this mechanism fails, as $F_2^{(\gamma)}(t)$ and $F_4^{(\gamma)}(t)$ saturate to nonzero values.

To further investigate the emergence of asymptotic freeness in the ergodic phase, it is necessary to check whether higher-order OTOCs also decay to zero. As illustrated in Fig.\,\ref{fig:F2n}, for a fixed $\gamma$, higher-order OTOCs indeed decay to zero, with the decay becoming increasingly rapid for larger values of $n$.

\begin{figure}[t]
\hspace*{-0.2 cm}
\includegraphics[width=1\textwidth]{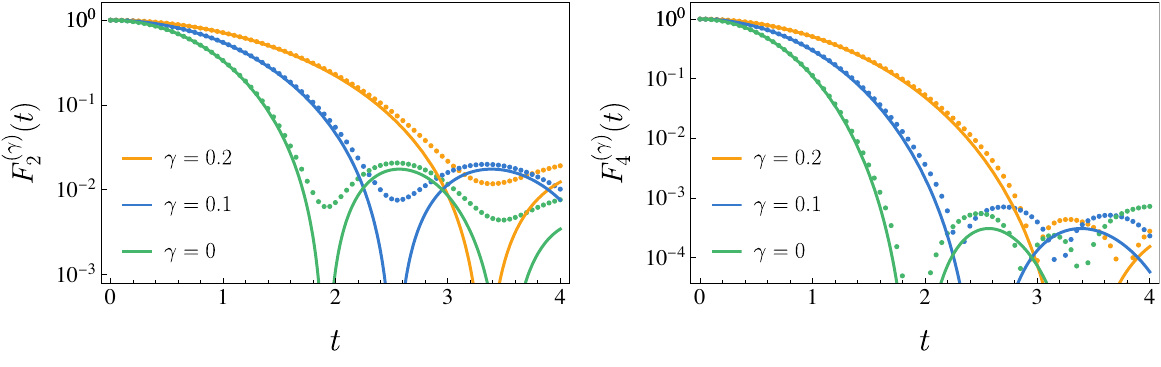}
\caption{Log-scale comparison of the numerical and analytical results of the two-point functions $F_2^{(\gamma)}(t)$ with $i=j=1$ (left) and four-point OTOCs $F_4^{(\gamma)}(t)$ with $i=1$, $j=L$ (right). Each panel includes three curves for $\gamma=0$ (green), $\gamma=0.1$ (blue), and $\gamma=0.2$ (orange). Dots indicate numerical results, while solid lines represent the analytic predictions from Eqs.\,\eqref{eq:F2_analytic} (left panel) and \eqref{eq:F4_analytic} (right panel). Here, we set $L=8$ and average over 100 realizations of the Hamiltonian.} \label{fig:LogF2andF4}
\end{figure}

Finally, in Fig.\,\ref{fig:LogF2andF4}, we compare the numerical results for $F_2^{(\gamma)}(t)$ and $F_4^{(\gamma)}(t)$ at small values of $\gamma$ and $s=1/2$ with the analytical expressions \eqref{eq:F2_analytic} and \eqref{eq:F4_analytic}. The analytical formulas accurately reproduce the data at early times and deviate only when the correlation functions approach very small values. This discrepancy arises because we are working with $L=8$, corresponding to a Hilbert space dimension $N=(2s+1)^L=256$. For such a relatively small Hilbert space, the correlation functions do not decay exactly to zero but instead exhibit residual fluctuations, resulting from the finite-size effects, that decrease as $L$ increases. By contrast, our analytical formulas are strictly valid in the limit $N \to \infty$. The resulting mismatch, though small, is visible in the log-scale plots, where deviations are more easily amplified.

\subsection{Cumulative OTOCs}

%\HC{I will add a few paragraphs to state how I would phrase this section. Please feel free to change the notation or modify the text. My logic is the following: we start with a general statement about the cumulative and weighted sum. Then we consider squared otocs for generic weights. Then we choose $1/n^2$ and show the results. Finally, we choose $f(n)=1/n$ and mention the connection with Shreya.}

As observed in the previous section, higher $n$-point functions exhibit a fast decay in the ergodic regime. Although this behavior persists in the fractal phase, such decay is significantly slower compared with the ergodic phase. Conversely, in the localized regime, these functions seem to retain their initial values and remain non-decaying for large time scales. Given that the decay of all $n$-point functions signals the emergence of freeness, we can characterize this emergence by considering the sum of such $n$-point functions. Such a sum would characterize the cumulative correlation between the operators. If $f(n)$ is a positive function of $n\in \mathbb{N}$ and $g(x)$ is a positive and continuous function of $x\in \mathbb{R}$, then we consider the partial sum of $p$ weighted and positive combinations of $n$-point functions $\mathcal{F}_p^{(\gamma)}(f,g\,;t)$ given by
\begin{align}
    \mathcal{F}_p^{(\gamma)}(f,g\,;t) := \sum_{n = 1}^{p} f(n) \, 
    g(F_{2n}^{(\gamma)} (t))\,, \label{sumogen}
\end{align}
where $F_{2n}^{(\gamma)} (t)$ are given by~\eqref{eq:npointfunctions}. Here, the role of $f(n)$ is to provide a specific weight to each of the functions $ g(F_{2n}^{(\gamma)} (t))$ in the partial sum, while $p$ represents the maximum number of $n$-point functions considered, and taking the limit $p\rightarrow \infty$ incorporates all sums of higher $n$-point functions. Note that since $g$ is a positive function, $\mathcal{F}_p^{(\gamma)}(f,g\,;t)$ will be zero if and only if the individual $g(F_{2n}^{(\gamma)}(t))$ vanish. Thus, this allows us to characterize the vanishing of all $n$-point functions using a single quantity.

The simplest positive function $g$ that we can consider is $g(x)=x^2$, leading to
\begin{align}
    \mathcal{F}_p^{(\gamma)}(f\,;t) := \sum_{n = 1}^{p} f(n) 
    (F_{2n}^{(\gamma)} (t))^2=\sum_{n = 1}^{p} f(n) \langle (S_1(0) S_1(t))^n \rangle^2 \,. \label{sumosquared}
\end{align}
The square is in a similar spirit to Vallini and Pappalardi \cite{Vallini:2024bwp}, where $g(x) = \log|x|$, without the cumulative sum. On the other hand, the sum in \eqref{sumosquared} thus corresponds to the cumulative contribution of weighted $n$-point squared OTOCs. At the same time, there is a large degree of freedom in choosing the weight $f(n)$. One way to choose it would be to fix the limiting value of the cumulative sum at $t=0$. In such a case, and for spin operators in the $p\rightarrow \infty$ limit, we have
\begin{align}
    \mathcal{F}_{\infty}^{(\gamma)}(f\,;0) := \sum_{n = 1}^{\infty} f(n)\,, \label{sumosquaredzerot}
\end{align}
since $\langle (S_1(0) S_1(0))^n \rangle^2=1$. One criterion for choosing $f(n)$ would be to ensure the convergence of ~\eqref{sumosquaredzerot}. Choosing, for example, $f(n)=n^{-k}$ for $\mathbb{Z}\ni k>1$ leads to $ \mathcal{F}_{\infty}^{(\gamma)}(0)=\zeta(k)$, where $\zeta(k)$ is the Riemann zeta function. Choosing $f(n)=n^{-1}$ also provides an alternative weight, which gives a finite result for finite $p$ (the Harmonic number $H_{p}$) but which otherwise diverges in the $p\rightarrow \infty$ limit.\footnote{We remark that such a choice is related to the convention of an independent study of operator spreading using free probability theory~\cite{Shreyatalk}.}

\begin{figure}[t]
\centering
\includegraphics[width=0.94\textwidth]{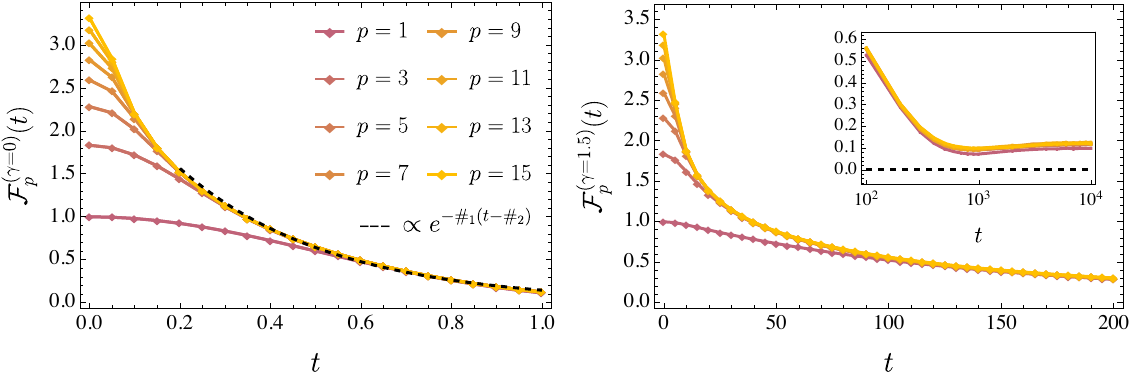}
\caption{Left: The cumulative OTOC \eqref{sumosquared} for the ergodic phase, keeping up to $p$-th order. Here we choose $f(n) = 1/n$ as the weight factor in \eqref{sumosquared}. The black dashed curve is the exponential decay obtained by summing up an infinite number of terms, given by Eq.\,\eqref{sumosquaredexpdecaylarget}. Right: The behavior of OTOC sum in the fractal phase, with an inset showing the late-time behavior. The parameters are $s = 1/2$, $L =8$ and $50$ ensemble averages of RP Hamiltonian is considered.} \label{fig:plotallgamma}
\end{figure}

\begin{figure}[t]
\centering
\includegraphics[width=0.94\textwidth]{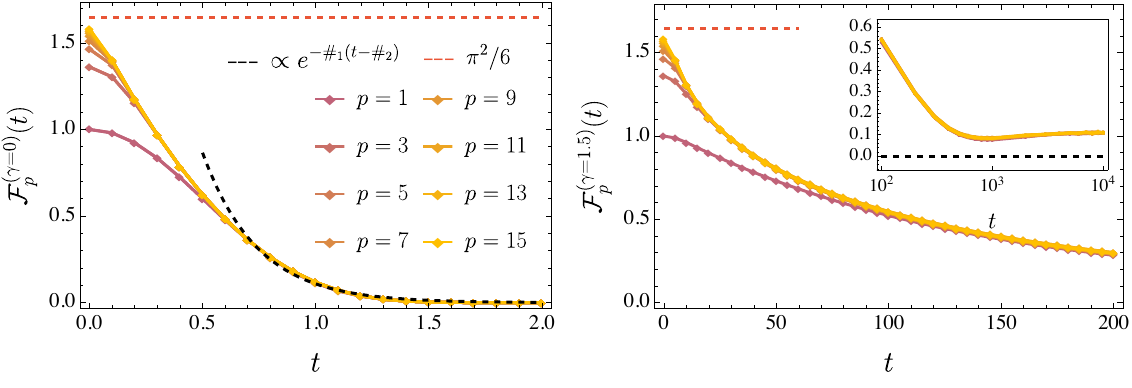}
\caption{Left: The cumulative OTOC \eqref{sumosquared} for the ergodic phase, keeping up to $p$-th order considering $f(n) = 1/n^2$. The saturation for the infinite sum is given by $\zeta(2)=\pi^2/6$. Right: The behavior of OTOC sum in the fractal phase, with an inset showing the late-time behavior. The parameters are $s = 1/2$, $L =8$ and $50$ ensemble averages of RP Hamiltonian is considered.} \label{fig:plotallgamma1bynq}
\end{figure}

Figures \ref{fig:plotallgamma} and~\ref{fig:plotallgamma1bynq} illustrate the behaviour of the cumulative partial sum $\mathcal{F}_p^{(\gamma)}(t):= \mathcal{F}_p^{(\gamma)}(f\,;t)$ for the choice of weights $f(n)=1/n$ and $f(n)=1/n^{2}$ respectively in the ergodic (left) and fractal (right) regimes, for $s=1/2$ and $L=8$. Since higher-point functions decay faster than lower-point functions \cite{Cotler:2017jue}, in practice, only a finite $p$ suffices to produce reasonable results.

Assuming an exponential decay of the $2$-point OTOCs \cite{Vallini:2024bwp}, the cumulative partial sum for a weight of the form $f(n)=n^{-k}$ for $\mathbb{Z}\ni k>0$ becomes
\begin{align}\label{sumosquaredexpdecay}
    \sum_{n = 1}^{p} \frac{1}{n^k}\, e^{- 2n  \chi t} =  -e^{-2(p+1)\chi t}\Phi(e^{-2\chi t},k,p+1)+\textrm{Li}_{k}(e^{-2\chi t})\,,
\end{align}
where $\Phi(a,b,c)$ represents the Lerch transcendent, Li$_{k}(z)$ is the polylogarithm function and $\chi$ denotes the exponential decay rate of $F_2^{(\gamma)}(t)$, which in general depends on the choice of operator. In the limit $p\rightarrow \infty$, only the term with polylogarithm remains, yielding
\begin{align}\label{sumosquaredexpdecaylarget}
    \sum_{n = 1}^{\infty} \frac{1}{n^k}\, e^{- 2n  \chi t} =  \textrm{Li}_{k}(e^{-2\chi t})\approx e^{-2\chi t}\,, \quad (\mathrm{large}~ t)\,.
\end{align}
This is indeed representative of the numerical results for the ergodic phase (left) in Figs.\,\ref{fig:plotallgamma} and~\ref{fig:plotallgamma1bynq}, where the horizontal red dashed line in the latter case corresponds precisely to $\zeta(2)=\pi^{2}/6$. The exponential decay is represented by the black dashed line in Figs.\,\ref{fig:plotallgamma} and~\ref{fig:plotallgamma1bynq} (left), demonstrating the eventual vanishing of the cumulative OTOC. The $\gamma= 0$ result primarily shows the result when the Hamiltonian is strictly controlled by the GOE matrices in the RP Hamiltonian. The same conclusion holds for other classes of Gaussian ensembles.

As an additional remark, it should also be noted that if we instead assume a slower decay, for example, a power-law behaviour of the form $ (F_{2n}^{(\gamma)} (t))^2=t^{-2\chi n}$, then the cumulative sum for $p\rightarrow \infty$ would lead to  $\mathcal{F}_{\infty}^{(\gamma)}(t)\approx t^{-2\chi}$ in the large time regime. This is consistent with findings of the decay of OTOCs and higher-point OTOCs in random matrix models \cite{Cotler:2017jue}.

In the fractal phase, the cumulative sum undergoes a decay, albeit at a much slower rate compared to the ergodic regime. As illustrated in the inset of Figs.\,\ref{fig:plotallgamma} and \ref{fig:plotallgamma1bynq} (right), at late times, the cumulative OTOC does not approach zero; rather, it decays to a small, non-zero value. This phenomenon reflects a form of ``memory'' retained within the operator statistics, which we will see in the subsequent section. In contrast, within the localized phase, the cumulative partial sum exhibits minimal decay, remaining almost constant at its initial value.

\section{Eigenvalue statistics of sum of operators from the free probability theory}\label{sec:operator_stat}

In this section, we investigate the eigenvalue statistics of the sum of two operators, one of which is time-evolved by a Hamiltonian drawn from the RP ensemble, and compare it with the prediction of free probability theory discussed in Sec. \ref{convolutions}. Specifically, our focus lies on the eigenvalue statistics of the following operator:
\begin{align}\label{sum_op}
    S(t) = S_1(0) + S_L (t) = S_1 (0) + U^{\dagger} (t) S_L(0) U(t)\,,~~~ U(t) = e^{-i H_{\mathrm{RP}} t}\,,
\end{align}
for specific values of the time parameter when the temporal evolution is generated by the RP model Hamiltonian, as defined in \eqref{HamRP}. Here $S_1(0)$ and $S_L(0)$ represent the spin operators at the initial time $t=0$, defined in \eqref{aop} and \eqref{bop}. The total sum of the operators $S(t)$ 
constructed through this formalism, encapsulates the effect of the RP Hamiltonian \eqref{HamRP} over time. Our primary objective is to systematically analyze the eigenvalue statistics of $S(t)$ at various time instants $t$. By varying $\gamma$, \emph{i.e.}, the parametric regime of the RP Hamiltonian, distinct phases (ergodic, fractal, or localized) or structures in the operator statistics are anticipated, potentially exhibiting novel characteristics inspired by predictions from free probability theory. 

\subsection{Choice of the initial operators} \label{sec:operators}

In the following analysis, we consider the operators $S_1(0)$ and $S_L(0)$ to be spin operators with spin $s$. This formulation generalizes the construction and enables comparisons against free probability predictions for various spin values, as obtained through analytical methods outlined in Section~\ref{sec:freeprobability}. The operators $S_1(0)$ and $S_L(0)$ are defined as tensor products of spin operators as in \eqref{aop}-\eqref{bop}.

\begin{figure}[t]
\centering
\includegraphics[width=0.92\textwidth]{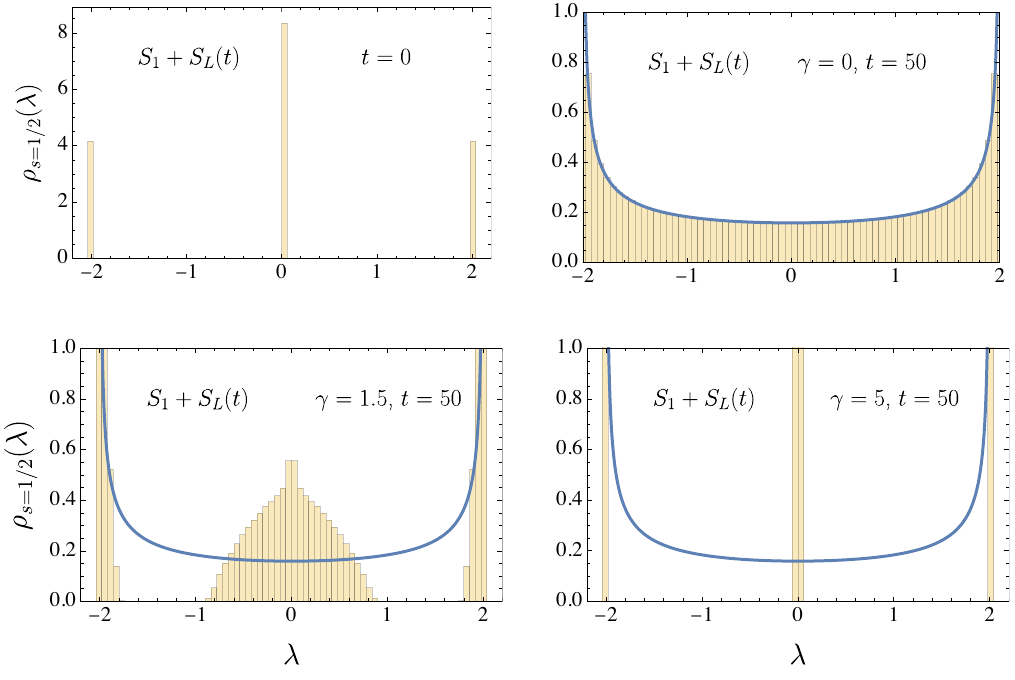}
\caption{The statistics of the eigenvalues of the operator \eqref{sum_op} with $s = 1/2$ at the initial time $t=0$ (top, left) and late times $t=50$ in the ergodic (top, right), fractal (bottom, left), and localized (bottom, right) regimes. We choose $L = 8$ (\emph{i.e.}, dimension of the Hamiltonian $N = (2s+1)^L = 256$) with $5000$ ensemble averages.  The solid blue lines describe the arsine distribution \eqref{arcsinedist} \emph{i.e.}, Fig.\,\ref{fig:allanalplot} (a).} \label{fig:opestatspinhalfall}
\end{figure}

We analyze three distinct spin values $s = \{1/2, 1, 3/2\}$, each paired with a corresponding lattice size $L = \{8,6, 5\}$. Using the Frobenius norm discussed earlier, the eigenvalues of the operators $S_1$ and $S_L$ are determined as follows: 

\begin{table}[h]
\centering
\begin{tabular}{c|c|c}
\textbf{Spin} \(s\) & \textbf{Operators} & \textbf{Eigenvalues} \\
\hline
\( 1/2 \) & \( S_1, S_L \) & \( \{1, -1\} \) \\
1                & \( S_1, S_L \) & \( \{1, 0, -1\} \) \\
\( 3/2 \) & \( S_1, S_L \) & \( \{3, 1, -1, -3\} \) \\
\end{tabular}
\caption{Eigenvalues of the operators $S_i$, with $i=1,L$ for different spin values.}
\label{tab:eigenvalues}
\end{table}

This is a convention that we follow throughout our discussion. More generally, we take any spin $s$ of our choice, with an $(2s+1)^{L-1}$ equal number of degenerate and distinct eigenvalues ranging from $-\epsilon s$ to $\epsilon s$ in steps of $\epsilon$, where $\epsilon = 2$ for half-integrer spins and $\epsilon = 1$ for integrer spins, as introduced in \eqref{aop}-\eqref{bop}. It is worth noting that alternative constructions of the higher-spin operators are possible. These constructions may lead to different commutation relations, such as: $[X_j^{(s)}, Y_k^{(s)}]= i n_s \delta_{jk} \, Z_j^{(s)}$, where $n_s = 1/s$ with the additive relations $X_i^{(s)} X_i^{(s)} + Y_i^{(s)} Y_i^{(s)} + Z_i^{(s)} Z_i^{(s)} = \big(\frac{s+1}{s}\big) \,I_{N}$, as discussed in Refs.\,\cite{Xu:2019lhc, Bhattacharjee:2022vlt}. Another possible formulation is: $[X_j^{(s)}, Y_k^{(s)}]= i \delta_{jk} \, Z_j^{(s)}$ with $X_i^{(s)} X_i^{(s)} + Y_i^{(s)} Y_i^{(s)} + Z_i^{(s)} Z_i^{(s)} = s(s+1) \,I_{N}$, as explored in Ref.\,\cite{Yin:2020oze}. Despite these differing constructions, it is straightforward to demonstrate that they are related through appropriate scalings of the spin operators. For instance: $X_j^{(s)} \rightarrow \sqrt{\frac{s+1}{3s}}X_j^{(s)}$ (similar for $Y_j^{(s)}$ and $Z_j^{(s)}$) for the first case \cite{Xu:2019lhc, Bhattacharjee:2022vlt} and $X_j^{(s)} \rightarrow \sqrt{\frac{s(s+1)}{3}}X_j^{(s)}$ (similar for $Y_j^{(s)}$ and $Z_j^{(s)}$) for the second case \cite{Yin:2020oze}. These scalings effectively normalize the eigenvalues to align with the conventions outlined in Table \ref{tab:eigenvalues}. Consequently, regardless of the specific construction chosen, the resulting operator statistics remain consistent and yield equivalent results.

\subsection{Operator statistics at different times}

Figure \ref{fig:opestatspinhalfall} illustrates the eigenvalue statistics of the operator \eqref{sum_op} with $s = 1/2$ across three distinct phases of the system: the ergodic phase $\gamma = 0$, the fractal phase $\gamma = 1.5$, and the localized phase $\gamma = 5$. These specific values of $\gamma$ are selected to ensure consistency with the eigenvalue statistics derived for the respective dimensions of the  Hamiltonian; see Fig.\,\ref{fig:eigstatRP}.

At the initial time $t=0$, the operator statistics are independent of the Hamiltonian and thus follow the \emph{multinomial distribution}, depending on their number of distinct eigenvalues (see Table~\ref{tab:eigenvalues}). This is what is expected from the  \emph{classical convolution} of the distributions of the eigenvalues of the individual operators. In the ergodic phase, the eigenvalue statistics of the operator exhibit convergence toward the arcsine distribution at the chosen timestamp, as predicted by free probability theory \cite{Camargo:2025zxr}. This behavior reflects the expected random matrix-like dynamics within the ergodic regime. In contrast, the eigenvalue distributions in the fractal and localized phases deviate significantly from the arcsine distribution at the same timestamp. These deviations highlight the role of different nature of the quantum dynamics in the fractal and localized regimes, and we can speculate that quantum dynamics in these regimes of the Hamiltonian do not drive the operators (here $S_1(0)$ and $S_L(0)$) towards freeness (see also the discussion on the late time behavior of these statistics below). 

\begin{figure}[t]
\centering
\includegraphics[width=0.92\textwidth]{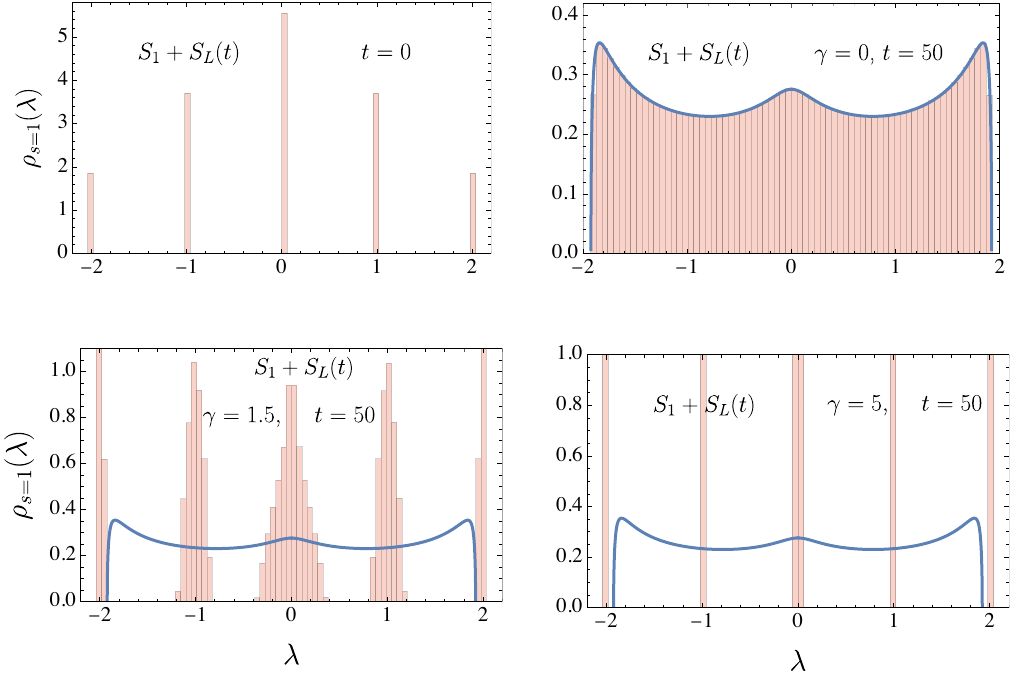}
\caption{The statistics of the eigenvalues of the operator \eqref{sum_op} with $s = 1$ at the initial time $t=0$ (top, left) and late times $t=50$ in the ergodic (top, right), fractal (bottom, left), and localized (bottom, right) regimes. We choose $L = 6$ (\emph{i.e.}, dimension of the Hamiltonian $N = (2s+1)^L = 729$) and average over $1000$ independent realizations.  The solid blue lines describe the distribution predicted by the free probability Eq.\,\eqref{eq: sumdis_spin1}, \emph{i.e.}, Fig.\,\ref{fig:allanalplot} (b).} \label{fig:opestatspin1all}
\end{figure}

\begin{figure}[t]
\centering
\includegraphics[width=0.92\textwidth]{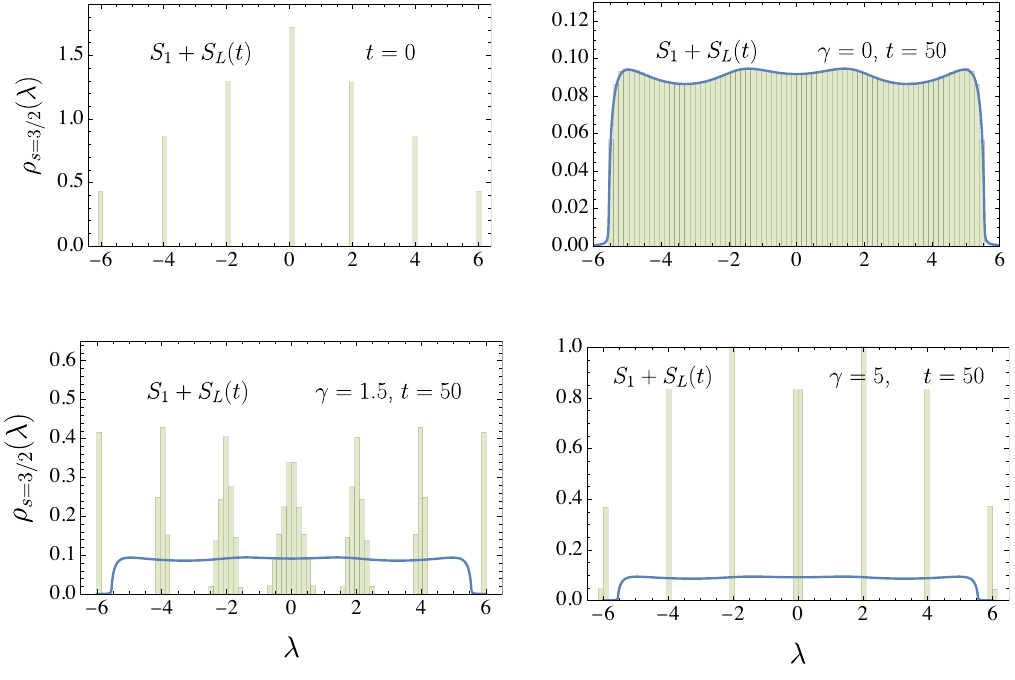}
\caption{The statistics of the eigenvalues of the operator \eqref{sum_op} with $s = 3/2$ at the initial time $t=0$ (top, left) and late times $t=50$ in the ergodic (top, right), fractal (bottom, left), and localized (bottom, right) regimes. We choose $L = 5$ (\emph{i.e.}, dimension of the Hamiltonian $N = (2s+1)^L = 1024$) and average over $500$ independent realizations of the Hamiltonian.  The solid blue lines describe the distribution predicted by the free probability, \emph{i.e.}, Fig.\,\ref{fig:allanalplot} (c).} \label{fig:opestatspin3by2all}
\end{figure}

The same computations can be extended to spin $s = 1$ and spin $s= 3/2$ operators given in \eqref{aop}-\eqref{bop}. The resulting eigenvalue statistics are illustrated in Fig.\,\ref{fig:opestatspin1all} and Fig.\,\ref{fig:opestatspin3by2all}, corresponding to $s=1$ and $s=3/2$ respectively. The analytical expression for the distribution predicted by the free probability theory for $s=1$  is given in Eq.\,\eqref{eq: sumdis_spin1}, derived previously in  \cite{Camargo:2025zxr}, while for $s=3/2$ the exact analytical expression of the free probability prediction is not known. The numerical findings for $s=1$ and $s=3/2$ remain consistent with the conclusions derived for the $s=1/2$.

\begin{figure}[t]
   \centering
\includegraphics[width=0.9\textwidth]{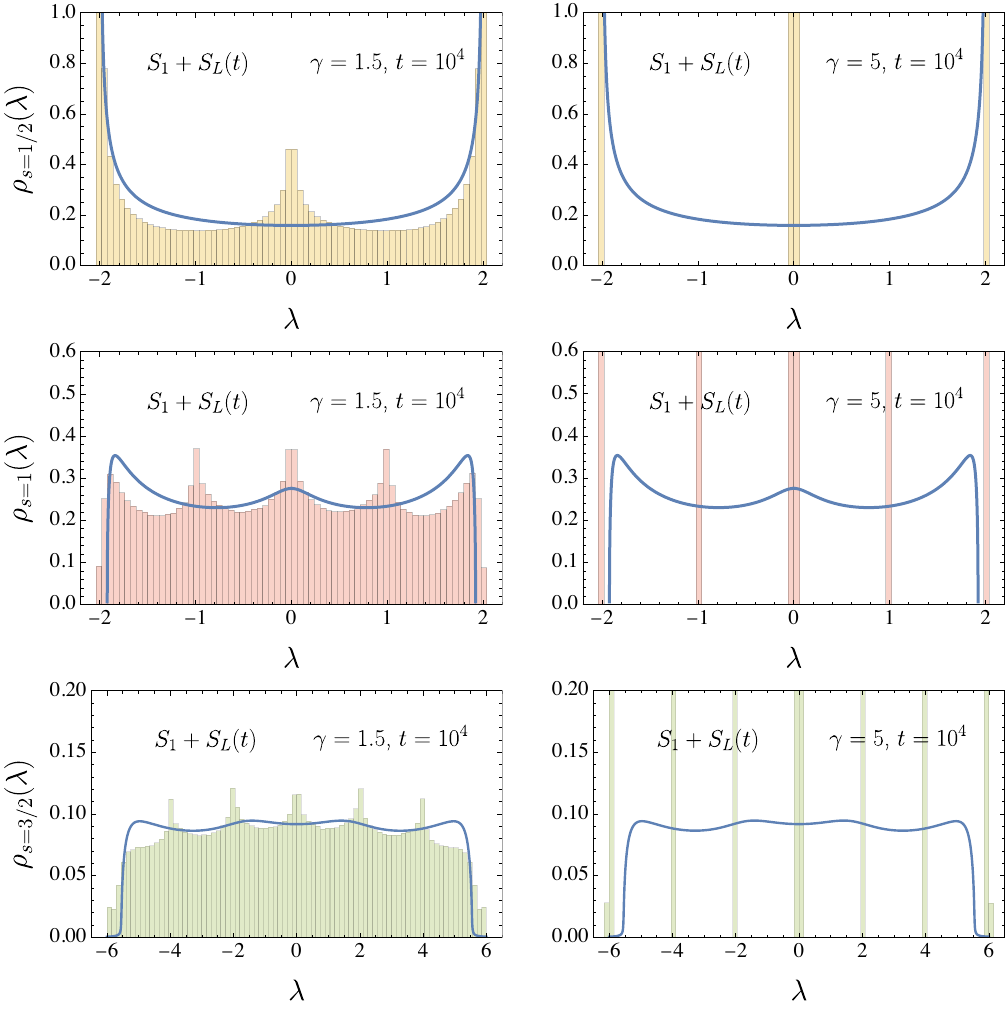}
\caption{The statistics of the eigenvalues of the operator \eqref{sum_op} with $s = 1/2$ (top panel),  $s = 1$ (middle panel), and  $s = 3/2$ (bottom panel) at the very late times $t \sim O(10^4)$ in the fractal (left panels) and the localized (right panels) phases. The solid lines in all panels describe the distribution predicted by the free probability. The system parameters are the same as Fig.\,\ref{fig:opestatspinhalfall} -- Fig.\,\ref{fig:opestatspin3by2all}. In the fractal regime, the operator statistics converge to a specific distribution that deviates from the free probability prediction. Distinct spikes are observed at the same points as those of at $t=0$, indicating that the time-evolved operator retain some `memory' of the initial distribution. Conversely, in the localized regime, the operator statistics remain markedly different from the results predicted by free probability, indicating the persistence of strong localization effects.} \label{fig:verylateopestatall}
\end{figure}

A natural question one can now ask is the following: do the eigenvalue statistics in the fractal and localized phases eventually approach the free probability prediction at sufficiently late times?\footnote{The specific shape of the distribution might hold its own intrinsic interest; however, our primary focus is on whether it converges to the free probability distribution. We thank Kohei Kawabata and Hal Tasaki for discussions on this point.} The results presented in Fig.\,\ref{fig:verylateopestatall} suggest that they do not. At very late times $t \sim O(10^4)$, for all the spins, the operator eigenvalue distribution in the fractal phase stabilizes into a distinct form, which remains fundamentally different from the free probability prediction. However, there are several spikes observed, exactly at the same point, keeping \emph{a memory} of the initial distribution at $t = 0$.  This indicates that the fractal phase sustains its intermediate, non-ergodic characteristics even at prolonged timescales. While we have considered two separate operators in the operator statistics, this effect of memory arises due to the non-vanishing contribution of the cumulative OTOC, as we observed in Fig.\,\ref{fig:plotallgamma} (right) and Fig.\,\ref{fig:plotallgamma1bynq} (right).

In the localized phase, the eigenvalue statistics display an even more striking behavior. Specifically, the operator statistics remain almost unchanged from their initial configuration $t = 0$, signaling a complete absence of dynamical evolution. This phenomenon aligns with the hallmark of localization, where the degrees of freedom of the system are effectively frozen, preventing significant operator spreading.

In summary, we conclude that in the ergodic phase, the \emph{classical convolution} of two commuting operators evolves into the \emph{free convolution} at late times. In contrast, in the fractal and localized phases, this transition does not occur.

\subsection{Analysis of the \emph{free time}}\label{time_scale}

In the preceding sections, we demonstrated that at late times, in the ergodic phase, the RP model Hamiltonian exhibits operator statistics consistent with predictions from free probability theory. However, the associated timescale predictions have thus far remained predominantly qualitative rather than quantitative. In this section, we aim to bridge this gap by undertaking a rigorous quantification of the timescales. Our approach incorporates a range of methodologies, including the use of distance metrics and advanced statistical techniques like hypothesis testing.

We begin with a two-distance measure, which computes the magnitude of the difference between the histogram data points and the corresponding analytical values. First consider the following function \cite{Camargo:2025zxr},
\begin{align}
    \chi^2 (t) = \sum_{i \neq \mathrm{bdy}} \bigg(\frac{\rho(\lambda_i,t) - \rho_{\mathrm{free}} (\lambda_i)}{\rho_{\mathrm{free}} (\lambda_i)} \bigg)^2\,, \label{chifunc}
\end{align}
where $\rho(\lambda_i,t)$ is the operator density at time $t$ and $\rho_{\mathrm{free}}(\lambda_i)$ denotes the eigenvalue density obtained from the free probability prediction. Here $\lambda_i$ is the $i$-th bin interval, which sums over the total number of full intervals of the operator statistics. It is important to note that this quantity is slightly different from other well-known \emph{Goodness-of-Fit tests} \cite{corder2014nonparametric, deshpande2017nonparametric}, such as Pearson's $\chi^2$ test, by its tailored focus on operator density deviations.

Another well-known information theoretic distance measure is the Kullback--Leibler (KL) divergence \cite{amari2000methods}, also known as the relative entropy in the context when  $\rho(\lambda) $ is the density matrix of some quantum mechanical system. Since we want to quantify the distance between a numerically obtained distribution and the distribution predicted by the free probability,  here we define the KL divergence as follows
\begin{align}
   \mathcal{D}_{\mathrm{KL}}(t) \equiv \mathcal{D}_{\mathrm{KL}}(\rho(\lambda_i,t)||\rho_{\mathrm{free}} (\lambda_i)) := \sum_{i \neq \mathrm{bdy}} \rho(\lambda_i,t) \bigg|\log \bigg[\frac{\rho(\lambda_i,t)}{\rho_{\mathrm{free}} (\lambda_i)}\bigg]\bigg|\,. \label{KLdiv}
\end{align}
In this definition, we have taken the absolute value of the term involving the logarithm to ensure that only the relative magnitude ratio of the two corresponding points of the histograms is important. However, note that in the usual definition of KL divergence, only the ratio $\rho(\lambda_i,t)/\rho_{\mathrm{free}} (\lambda_i)$ appears, not its absolute value. At earlier times when $\rho(\lambda_i,t) = 0$ for some particular $\lambda_i$, the above expression needs to be taken as the limit case such that $\lim_{x \rightarrow 0} x \log x = 0$.

\begin{figure}[t]
\centering
\includegraphics[width=0.92\textwidth]{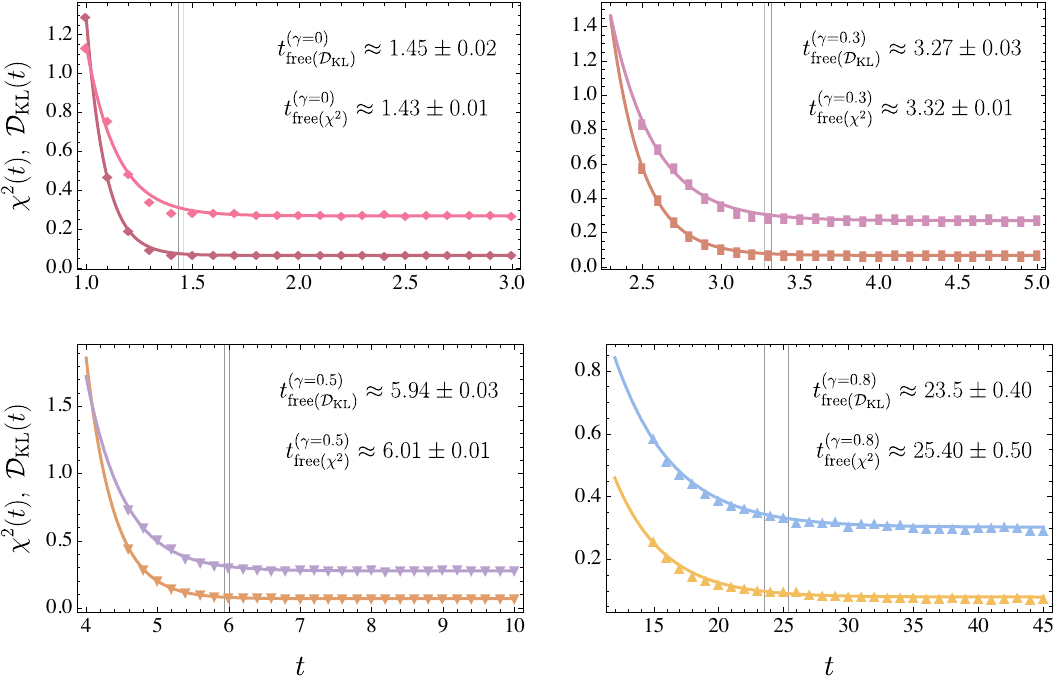}
\caption{The behavior of the $\chi^2$ function and the KL divergence with time for $\gamma  = 0$ (top, left), $\gamma  = 0.3$ (top, right), $\gamma  = 0.5$ (bottom, left), and $\gamma  = 0.8$ (bottom, right).  In each panel, the data points for $\chi^2$ lie above the data points for the KL divergence. We choose $s = 1/2$, $L = 8$, and take averages over $500$ realizations.}  \label{fig:chisqvstplot}
\end{figure}

Figure \ref{fig:chisqvstplot} shows the temporal evolution of the $\chi^2$ function and the KL divergence for four representative values of $\gamma$. The solid gray lines mark the corresponding \emph{free times}, obtained in each case by fitting the data to the following exponentially decaying function:
\begin{align}
    \mathcal{F}_{\mathrm{fit}}(t) := \mathcal{B} \,e^{-(t-t_0)/t_d} + \mathcal{B}\,,~~~~~~~  \mathcal{F}_{\mathrm{fit}} = \chi^2_{\mathrm{fit}}~~ \mathrm{or} ~~(\mathcal{D}_{\mathrm{KL}})_{\mathrm{fit}}\,,
\end{align}
where $\mathcal{B}$, $t_0$, and $t_d$ are fitting parameters yet to be determined. As $t \rightarrow \infty$, the function $\mathcal{F}_{\mathrm{fit}}(t) \rightarrow \mathcal{B}$, signifying that $\mathcal{B}$ represents the saturation value at late times.

The \emph{free time}, denoted as  $t_{\mathrm{free}}$, represents the timescale that satisfies the following condition:
\begin{align}
    t_{\mathrm{free}}:= t_0 + n \,t_d\,, ~~ n = 1,2,\cdots,~~~ \Rightarrow ~~~ \mathcal{F}_{\mathrm{fit}}(t_{\mathrm{free}}) = \mathcal{B} \left(\frac{1}{e^n} + 1\right)\,. \label{free}
\end{align}
To estimate $t_{\mathrm{free}}$, we identify the timescale at which the difference between $\mathcal{F}_{\mathrm{fit}}(t)$ and its saturation value $\mathcal{B}$ satisfies:
\begin{align}
    |\mathcal{F}_{\mathrm{fit}}(t) - \mathcal{B}| = \frac{\mathcal{B}}{e^n} \leq \delta\,,
\end{align}
where $\delta$ is a prescribed tolerance.  Determining $n$ to fulfill this condition and substituting it into \eqref{free}, we obtain the \emph{free time}. 

It is important to note that at late times, the saturation value of the KL divergence exceeds that of the $\chi^2$ function.\footnote{We thank Tanay Pathak for asking this question.} However, this does not imply that the KL divergence is inferior to the $\chi^2$ function. The distinction arises from the very nature of their definitions, as given in \eqref{chifunc} and \eqref{KLdiv}. To illustrate this, consider an instance when the operator statistics align with the predictions of free probability but exhibit small numerical fluctuations around the predicted value. Define $\rho_{\mathrm{free}}(\lambda_i) \simeq u$ and $\rho(\lambda_i,t) \simeq u + \epsilon$, where $u$ is constant and $\epsilon \ll u$ represents a minor fluctuation relative to $u$. Using these, the respective distance measures evaluate as:
\begin{align}
    \chi^2 &\simeq \bigg(\frac{(u + \epsilon) - u}{u} \bigg)^2 = \frac{\epsilon^2}{u^2}\,,\nonumber \\
    \mathcal{D}_{\mathrm{KL}} &\simeq (u+\epsilon) \bigg|\log\bigg[\frac{u+\epsilon}{u}\bigg]\bigg| = \epsilon + \frac{\epsilon^2}{u}\,.
\end{align}
Given that $\epsilon \ll u$, it follows that $\chi^2 \simeq O(\epsilon^2)$ at leading order, while $\mathcal{D}_{\mathrm{KL}} \simeq O(\epsilon)$. This establishes that by definition, the saturation value of $\chi^2$ will be lower than that of the KL divergence, justifying the results in Fig.\,\ref{fig:chisqvstplot}.

\begin{figure}[t]
\hspace*{-0.5 cm}
\begin{subfigure}[b]{0.5\textwidth}
\centering
\includegraphics[width=\textwidth]{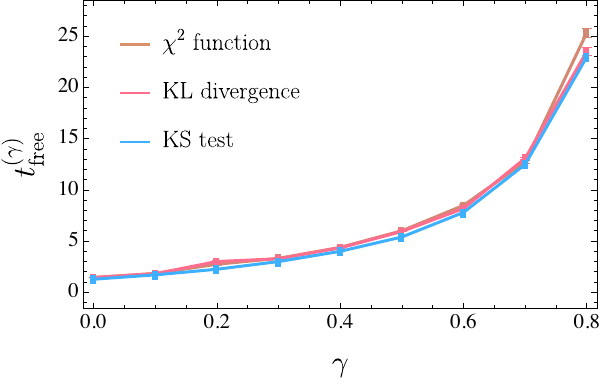}
\caption{}
\end{subfigure}
\hfil
\begin{subfigure}[b]{0.5\textwidth}
\centering
\includegraphics[width=\textwidth]{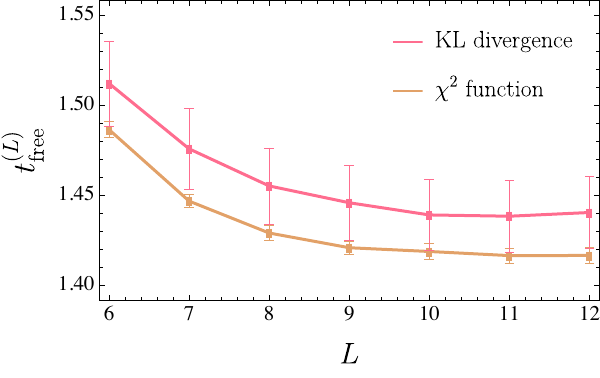}
\caption{}
\end{subfigure}
\caption{(a) The \emph{free time} (or the arcsine time for spin $1/2$) with the deformation $\gamma$, computed from the $\chi^2$ function (brown), KL divergence (red), and the Kolmogorov-Smirnov (KS) test (blue). (b) The finite-size scaling of \emph{free time} for a fixed $\gamma = 0$. The corresponding realizations are taken as $L = 6 \,(5000)$, $L=7 \,(3000)$, $L = 8\, (1000)$, $L = 9 \,(500)$, $L = 10 \,(200)$, $L = 11\, (100)$, and $L = 12\, (50)$. The dimension of the Hilbert space is $2^L$. The errors are computed by the standard error estimates in \emph{Mathematica}.}
\label{fig:freetimegammaplot}
\end{figure}

Figure \ref{fig:freetimegammaplot} (a) shows the behavior of the \emph{free time} with respect to $\gamma$ (brown for $\chi^2$ function and red for KL divergence). We chose $n = 2$ in \eqref{free} such that the error remains within $10\%$ of the saturation value. As evident, the time increases as $\gamma$ increases, \emph{i.e.}, the operator statistics lead to the free probability prediction at later times as the system moves away from the ergodic regime. The dependence of \emph{free time} on $L$ (the system size being $2^L$) from $L = 6$ to $L=12$ is shown in Fig.\,\ref{fig:freetimegammaplot} (b) for both distance measures. The error estimates are computed by the standard error estimates in \emph{Mathematica}. The errors for the $\chi^2$ function and KL divergence are of $O(10^{-3})$ and $O(10^{-2})$ respectively.

The final method we consider is the Kolmogorov-Smirnov (KS) test~\cite{kolmogorov1933,smirnov1939estimation,smirnov1948}, a non-parametric statistical test used to determine whether two one-dimensional probability distributions are identical. It is commonly employed to assess whether empirical data could plausibly originate from a given reference distribution. In the language of hypothesis testing, the assumption that both distributions are equal defines the null hypothesis, denoted by $H_0$, while the alternative hypothesis, $H_1$, corresponds to the case in which they differ~\cite{corder2014nonparametric}.

Let us consider an observed sample ${ \lambda_1, \lambda_2, \cdots,\lambda_N }$ consisting of $N$ data points. We aim to determine whether this sample follows a specific reference probability density function, denoted by $\text{PDF}_\text{ref}(\lambda)$. The KS test compares the empirical cumulative distribution function (CDF) of the sample with the CDF of the reference distribution. The latter is computed as
\begin{equation}
\text{CDF}_\text{ref}(\lambda)= \int_{-\infty}^{\lambda} \text{PDF}_\text{ref}(\lambda') \,d\lambda'\,.
\end{equation}
The empirical CDF is defined as the proportion of sample points less than or equal to a given value $\lambda$, and can be written as
\begin{equation}
\text{CDF}_N(\lambda)=\frac{1}{N}\sum_{i=1}^{N}\theta(\lambda-\lambda_i)\,,
\end{equation}
where $\theta$ is the Heaviside step function.

\begin{figure}[t]
\centering
\includegraphics[width=0.92\textwidth]{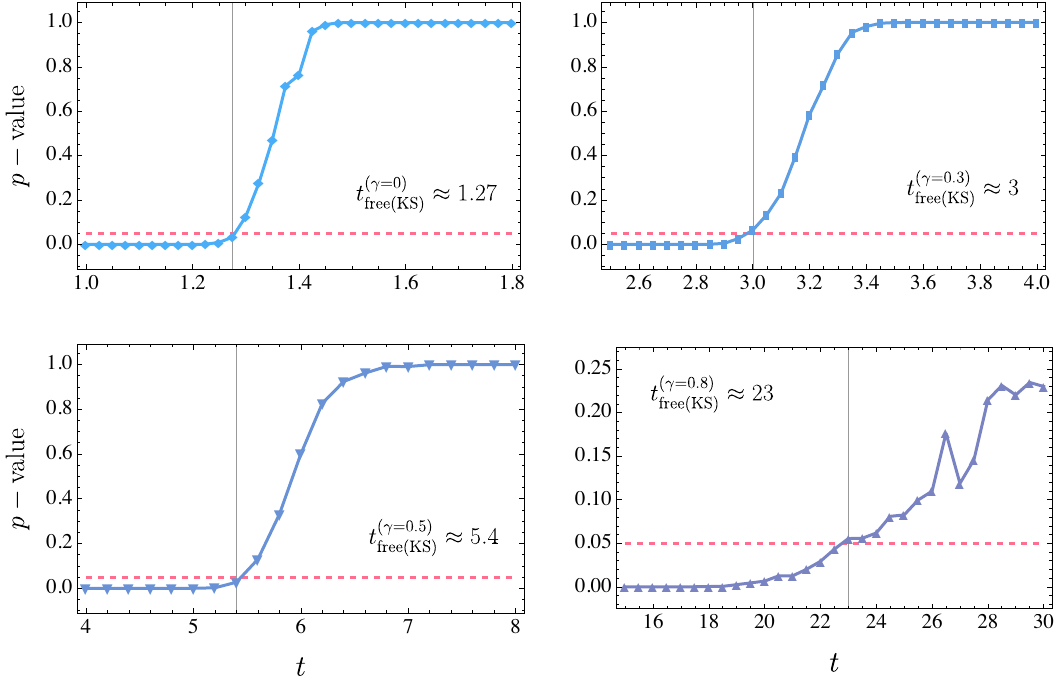}
\caption{The temporal evolution of $p$-value for the deformation $\gamma$ from the KS test. The red dashed line indicates the value of $p=0.05$, above which the null hypothesis is accepted, and the empirical distribution of the operator follows the free probability prediction. The grey dashed line indicates the \emph{free time}, where the curves intersect the red dashed line.} \label{fig:KSpvalvstplot}
\end{figure}

\begin{figure}
    \centering
    \includegraphics[width=0.55\linewidth]{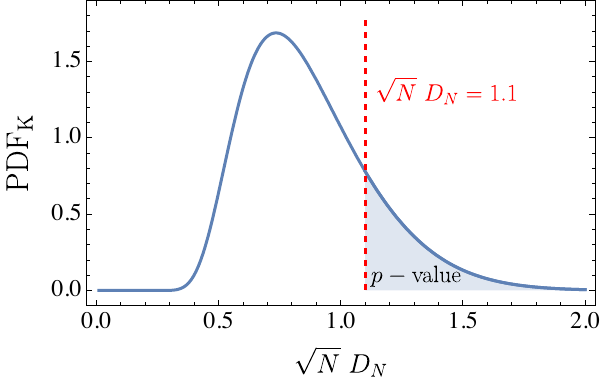}
    \caption{Kolmogorov probability density function (PDF). The shaded area illustrates the $p$-value associated with a particular value of \( x=\sqrt{N}D_N \). In hypothesis testing, a $p$-value below a chosen significance level -- commonly 0.05 -- indicates that the null hypothesis (that the two distributions are the same) can be rejected with high confidence. For the value \( \sqrt{N} D_N = 1.1 \) shown by the red dashed line, the $p$-value equals 0.178. In this case, the two distributions would be considered the same.}
    \label{fig:KolmogorovDist}
\end{figure}

The KS statistic is then defined as the maximum absolute difference between the empirical and reference CDFs across all $\lambda$ values~\cite{corder2014nonparametric}:
\begin{equation} \label{eq-Dn}
D_N = \sup_{\lambda} |\text{CDF}_N(\lambda)-\text{CDF}_\text{ref}(\lambda)|\,.
\end{equation}
Under the assumption that the null hypothesis is true, and in the limit of large $N$, the rescaled statistic $\sqrt{N} D_N$ converges in distribution to a random variable $x$ governed by the Kolmogorov distribution $\text{PDF}_{\mathrm{K}}$:
\begin{equation}
\lim_{N \rightarrow \infty} \sqrt{N} D_N \rightarrow x \sim \text{PDF}_{\mathrm{K}}(x)\,.
\end{equation}
The Kolmogorov probability density function is given by
\begin{equation}
    \text{PDF}_{\mathrm{K}}(x) = \frac{d}{dx} \text{CDF}_{\mathrm{K}}(x)= 8x \sum_{n=1}^{\infty} (-1)^{n-1} n^2 e^{-2n^2 x^2}\,.
\end{equation} 
The associated cumulative distribution function is given by~\cite{KStest}
\begin{equation}
\text{CDF}_{\mathrm{K}}(x)= \int_{-\infty}^{x} \text{PDF}_{\mathrm{K}}(x')\,dx' = 1 - 2 \sum_{n=1}^{\infty} (-1)^{n-1} e^{-2n^2x^2}\,,
\end{equation}
The procedure of the KS test involves computing the statistic $x=\sqrt{N} D_N$ and evaluating whether it is significantly large. This is done by evaluating the associated $p$-value $= 1-\text{CDF}_\text{K}(x)$, which is the probability of finding a larger value for the parameter $x=\sqrt{N} D_N$ given that the null hypothesis is true (the two distributions are the same). If the $p$-value is smaller than 0.05, one concludes that the parameter is too large and the null hypothesis is false, namely, the two distributions are not the same. Otherwise, if the $p$-value is larger than 0.05, one concludes that the null hypothesis is true and the two distributions are the same.\footnote{Given two distributions, the KS test can be easily implemented in \emph{Mathematica} by using the command \texttt{KolmogorovSmirnovTest}.} Figure \ref{fig:KolmogorovDist} illustrates the Kolmogorov PDF and the $p$-value associated with a particular value of $x=\sqrt{N} D_N$.

We apply the KS test to assess whether the empirical distributions for the eigenvalues of $A(0)+A(t)$ follow the prediction from free probability. In Fig.\,\ref{fig:KSpvalvstplot}, we show the $p$-value of the test as a function of time for different values of $\gamma$. The figure shows that at early times the $p$-value takes values smaller than 0.05, indicating that the empirical distribution and the free probability prediction are not the same. However, as time increases, the $p$-value increases, taking values larger than 0.05, indicating that the free probability prediction has emerged. The free time is computed as the time at which the $p$-value crosses the value 0.05.\footnote{We also checked with other tests such as Pearson's $\chi^2$ test and confirm the strikingly similar behavior of \emph{free time}, as in Fig.\,\ref{fig:freetimegammaplot}. }

\section{Conclusion and outlook} \label{concsec}

Within the framework of free probability theory, we have revisited the calculation of the DOS for the Rosenzweig–Porter (RP) random matrix ensemble and studied the spectral statistics of operators to understand the emergence of asymptotic freeness in different phases of this model. Since the RP model is defined as the sum of two independent matrices, one of which, in the standard version of this model, is drawn from the classical Gaussian random matrix ensembles, the two terms are free according to the free probability theory in the limit when the dimension of the matrices is sufficiently large. This enables us to use free probabilistic techniques to compute the DOS of this model. Using a general approach based on subordination functions, we derive an explicit expression for the DOS of the RP model in the fractal phase, reproducing previous results obtained via the replica technique~\cite{Venturelli:2022hka}. Our analysis focuses on the fractal phase, where we could write the DOS as perturbative corrections to the Gaussian profile characteristic of the localized phase. As illustrated in Fig.\,\ref{fig:DOSschematic}, the DOS evolves smoothly from a Wigner semicircle to a Gaussian distribution as the parameter $\gamma$ increases. In particular, during the transition from the ergodic to the fractal phase, the DOS resembles a perturbed Wigner semicircle. It would be interesting to extend the methodology developed here to compute the DOS in this intermediate regime. 

It has been recently proposed that chaotic dynamics drives the components of a sum of two operators, one of which is time-evolved, toward asymptotic freeness with respect to non-evolved ones \cite{Camargo:2025zxr, Chen:2024zfj}. However, the precise mechanisms underlying the emergence of freeness remain unclear. For instance, is it sufficient for the spectral statistics of the Hamiltonian to follow RMT? What role do the eigenvectors play in this process? To address these questions, we studied the emergence of free probability predictions for the spectrum of operators of the form $S_1 + S_L(t)$ in the RP model, an ensemble of random matrix Hamiltonians where the rotational invariance of the classical Gaussian ensembles is broken explicitly by introducing a parameter, and depending on the value of this parameter the energy eigenvectors show different types of localization properties. Our analysis reveals that the transition from classical convolution to free convolution manifests exclusively in the ergodic phase, where the eigenvectors are fully delocalized. In contrast, for neither the fractal phase nor the localized phase do the operator spectral statistics match the free convolution prediction.

Interestingly, in the fractal phase, the system exhibits partial signatures of freeness, yet retains memory of the spectrum at $t=0$. This memory is associated with the non-vanishing cumulative OTOCs at late times, which suggests that the emergence of freeness is not solely governed by spectral statistics, but also by the structure of the eigenvectors. While the fractal phase still displays level repulsion---as reflected in the average $r$-parameter---the corresponding eigenvectors are not fully delocalized. This points to a natural conclusion: the presence of level repulsion in the Hamiltonian alone is not sufficient for the emergence of freeness in operator statistics; extended, non-localized eigenvectors are also essential. It would be interesting to explore this direction further and test its validity in other models that exhibit fractal or multifractal phases \cite{Khaymovich:2021tkj}.

To substantiate the above conclusions, in Appendix \ref{app:DecorrEns} we study the eigenvalue statistics of the same time-evolved operators where the time evolution is now generated by a Hamiltonian drawn from a decorrelated version of the RP ensemble. The characteristic feature of the decorrelated ensembles (also known as the Poissonian ensembles in the literature \cite{Magan:2024nkr}) constructed from a given physical ensemble (or a quantum many-body system) is that a Hamiltonian drawn from it has the same eigenvector localization/delocalization/fractal property as that of the original ensemble, while, if there are correlations between the eigenvalues in the original ensemble they are removed by hand when constructing the decorrelated ensemble. Our results, presented in Appendix \ref{app:DecorrEns}, show that the decorrelated version 
of the RP ensemble gives rise to the same eigenvalue statistics of the time-evolved operators as those of the original RP model, once again indicating that complete delocalization of eigenvectors is sufficient 
for producing eigenvalue statistics compatible with free probability, 
not eigenvalue correlation.

In section \ref{time_scale}, we have utilized various measures to accurately characterize the timescale at which free probability predictions emerge, marking the onset of \emph{free time}. These methods include distance metrics like the $\chi^2$ function and KL divergence, alongside statistical techniques such as hypothesis testing. Notably, all methods yield consistent and reliable results. Additionally, the agreement between the KS test (based on the CDF) and both the KL divergence and $\chi^2$ test (based on the PDF) underscores the robustness of PDF-based methods in our analysis. This consistency confirms that the choice of histogram bin size does not affect our findings within the estimated error ranges, a conclusion supported by the considerable number of Hamiltonian ensembles in the operator statistics evaluation.

Several compelling directions emerge for future research. Notably, the sharp transition observed in the RP model manifests in the infinite-size limit, precisely the regime where predictions from free probability theory become exact. This alignment makes it particularly intriguing to attempt to identify the transition point within a free probability framework. A promising approach might involve fixing a sufficiently large time and analyzing the variation of the KL divergence with respect to $\gamma$, which measures the distance between the empirical distribution and the free probability distribution. Such an analysis could potentially reveal whether this distance provides a clear demarcation for the distinct transitions within the model. We expect to report on this in future work.

In the present study, our focus has been restricted to the statistical properties of eigenvalues of an operator of the form $S_1+S_L(t)$. However, the role of eigenvectors is equally crucial, especially in the context of thermalization. A natural extension would be to investigate the localization and delocalization behavior of the eigenvectors of the operator $S_1+S_L(t)$, which could shed light on the mechanisms underlying the ETH. Such an investigation would be especially illuminating when approached through the framework of free cumulants, potentially revealing interesting connections with recent developments in the field \cite{Pappalardi:2022aaz, Pappalardi:2023nsj, Fritzsch:2024qjn, Wang:2025mzl, Vallini:2024bwp, Fava:2023pac}, with our results on operator statistics. Especially, it would be an excellent direction to compare the \emph{free time} studied here with the \emph{timescale for the onset of freeness} introduced in Ref.\,\cite{Vallini:2024bwp}, generalizing to systems hosting multifractal states \cite{Khaymovich:2021tkj}.  Pursuing this direction would require the consideration of finite-temperature states, thereby necessitating a generalization of the current statistical framework. Addressing these questions will be the subject of future investigations.

\section{Acknowledgements}
We thank Neil Dowling, Yichao Fu, Koji Hashimoto, Kohei Kawabata, Ivan M. Khaymovich, Tokiro Numasawa, Silvia Pappalardi, Tanay Pathak, Hal Tasaki, and Masafumi Udagawa for fruitful discussions. We also thank Jonah Kudler-Flam for useful comments on the draft. P.N. acknowledges the hospitality of Gwangju Institute of Science and Technology (GIST), Gakushuin University, and the Institute of Solid State Physics (ISSP), University of Tokyo, during the final stages of the work. This work was supported by the Basic Science Research Program through the National Research Foundation of Korea (NRF) funded by the Ministry of Science, ICT \& Future Planning (NRF-2021R1A2C1006791), the framework of international cooperation program managed by the NRF of Korea (RS-2025-02307394), the Creation of the Quantum Information Science R\&D Ecosystem (Grant No. RS-2023-NR068116) through the National Research Foundation of Korea (NRF) funded by the Korean government (Ministry of Science and ICT), the Gwangju Institute of Science and Technology (GIST) research fund (Future leading Specialized Resarch Project, 2025) and the Al-based GIST Research Scientist Project grant funded by the GIST in 2025. This research was also supported by the Regional Innovation System \& Education (RISE) program through the (Gwangju RISE Center), funded by the Ministry of Education (MOE) and the (Gwangju Metropolitan City), Republic of Korea (2025-RISE-05-001). V.J. and H.C. are supported by the Basic Science Research Program through the National Research Foundation of Korea (NRF) funded by the Ministry of Education (NRF-2022R1I1A1A01070589, and RS-2023-00248186). The work of P.N. is supported by the JSPS Grant-in-Aid for Transformative Research Areas (A) ``Extreme Universe'' No. 21H05190.

\appendix

\section{Comparison of Haar random unitaries and exponential of Gaussian ensembles}
\label{app:ExpGauss}

\begin{figure}[t]
\hspace*{-0.6 cm}
\includegraphics[width=1.04\textwidth]{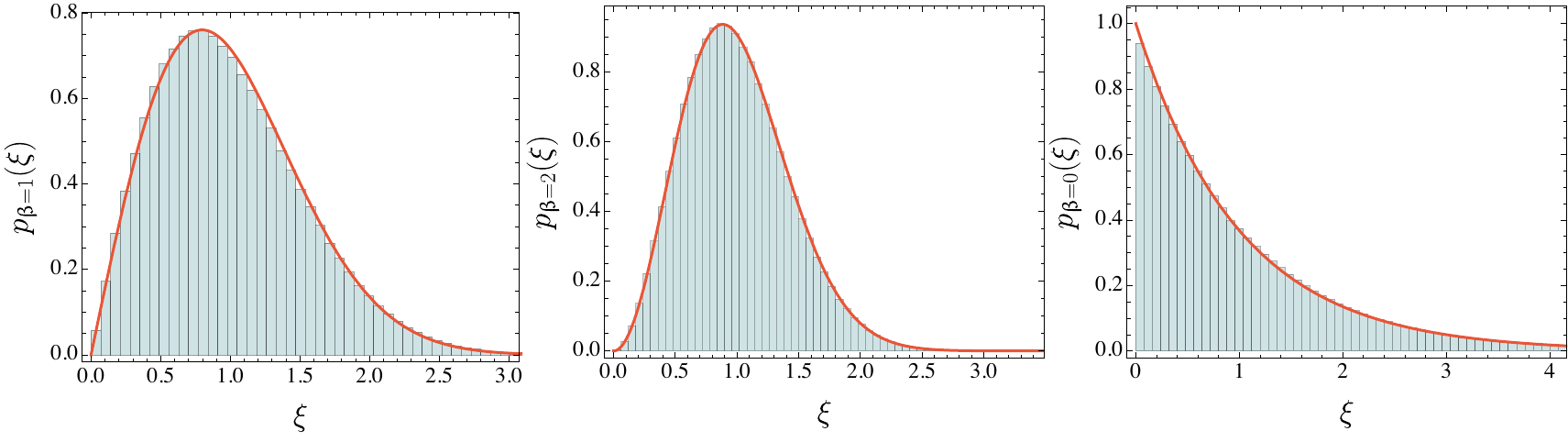}
\caption{Nearest-level spacing distribution for the unitaries $U_{\mathrm{COE}}$ (left), $U_{\mathrm{CUE}}$ (middle) and $U_{\mathrm{GXE}}$ (right), with $\mathrm{X} \in \{\mathrm{O}, \mathrm{U}\}$ (both produce identical plots) for $s = 1/2$. The red lines are the analytic results for $\upbeta = 1,2$ and $0$ respectively (from left to right). We choose $L = 8$ and average over $5000$ samples. For $U_{\mathrm{GXE}}$, we choose $t = 50$.} \label{sdiffhaar}
\end{figure}

\begin{figure}[t]
\hspace*{-0.6 cm}
\includegraphics[width=1.04\textwidth]{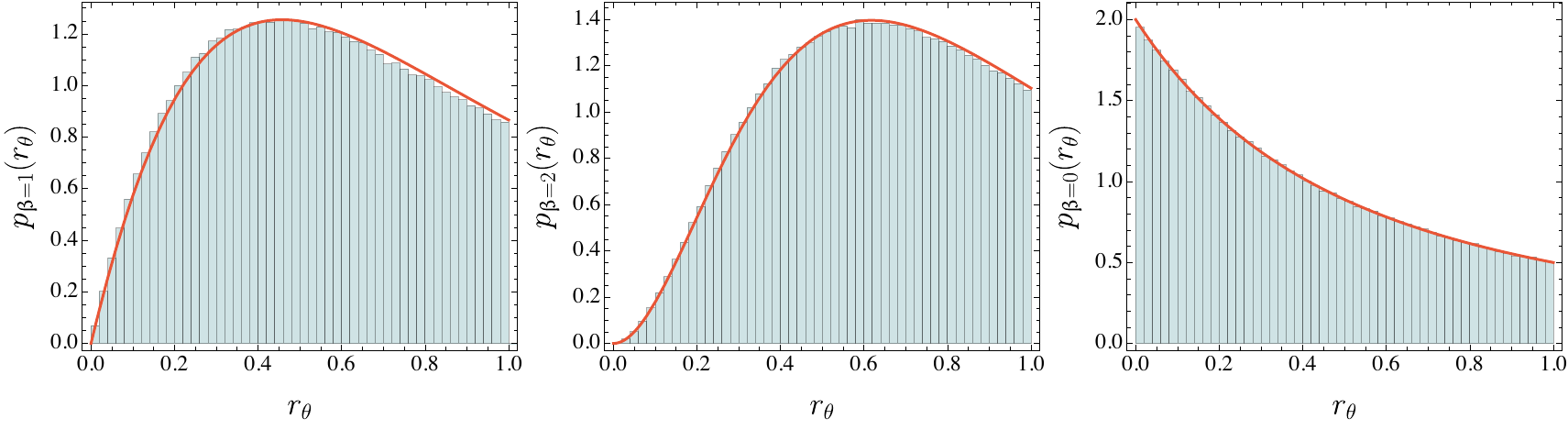}
\caption{Histogram of the $r$-ratio for the unitaries $U_{\mathrm{COE}}$ (left), $U_{\mathrm{CUE}}$ (middle) and $U_{\mathrm{GXE}}$ (right), with $\mathrm{X} \in \{\mathrm{O}, \mathrm{U}\}$ (both produce identical plots) for $s = 1/2$. The red lines are the analytic results for $\upbeta = 1,2$ and $0$ respectively (from left to right). We choose $L = 8$ and average over $5000$ samples. For $U_{\mathrm{GXE}}$, we choose $t = 50$.} \label{rratiohaar}
\end{figure}

In this appendix, we compare the asymptotic freeness governed by different $N \times N$ unitary ensembles. The unitary ensembles considered are listed below:
\begin{align}
    U_{\mathrm{CXE}}\,,  ~~~ U_{\mathrm{GXE}} = e^{-i H_{\mathrm{GXE}} t}\,, ~~~ \mathrm{X} \in \{\mathrm{O}, \mathrm{U}\}\,, \label{unitaries}
\end{align}
with $t \sim O(1)$. This list includes the corresponding orthogonal and unitary ensembles. The conclusions for the symplectic ensemble can also be generalized straightforwardly. Among the unitaries in \eqref{unitaries}, $U_{\mathrm{CXE}}$ with $\mathrm{X} \in \{\mathrm{O}, \mathrm{U}\}$ are the Haar random, while $U_{\mathrm{GXE}} = e^{-i H_{\mathrm{GXE}} t}$ with $t \sim O(1)$ are not. Following \cite{Cotler:2017jue}, we refer to the latter as an \emph{ensemble of Gaussian Hamiltonians}. This distinction can be easily verified by examining the spectrum of the phase of the eigenvalues of the corresponding unitaries. More specifically, for all cases, the eigenvalues are of the form $\lambda_k = r e^{i \theta_k}$, where $\theta_k = \mathrm{Arg}(\lambda_k) \in [-\pi, \pi]$ and $k = 1, \cdots, N$. Therefore, we consider the spacing distribution of  $\theta_k$, defined as $\xi_j = \frac{N}{2 \pi}(\theta_{j+1} - \theta_j)$ with $j = 1, \cdots, N-1$, and the $r$-ratio, defined in \eqref{rratioformula} for all unitaries \eqref{unitaries}. While the DOS $\rho(\theta) = 1/(2\pi)$ is uniform, the unitaries $U_{\mathrm{CXE}}$ exhibit level repulsion while $U_{\mathrm{GXE}}$ show behavior typical of an uncorrelated ensemble. The nearest-level spacing distribution $\xi_k$ and the $r$-ratio distribution are illustrated in Fig.\,\ref{sdiffhaar} and Fig.\,\ref{rratiohaar} respectively. The analytic results are matched with \eqref{eq:WD} and \eqref{pr} respectively.

\begin{figure}[t]
\centering
\includegraphics[width=0.92\textwidth]{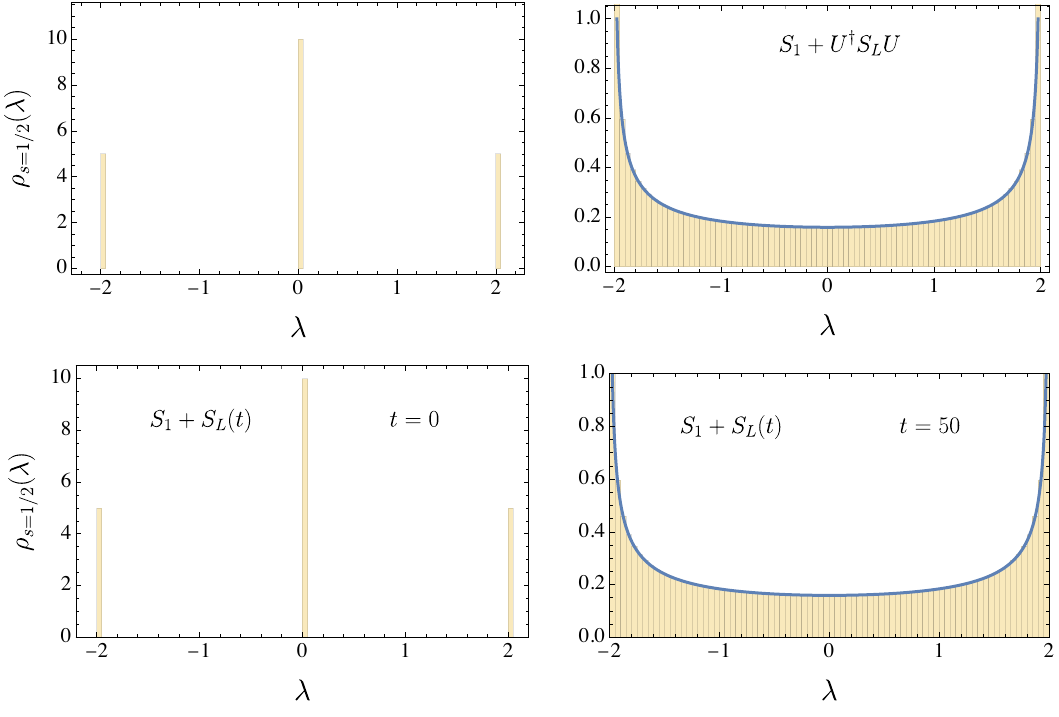}
\caption{The operator statistics governed by the unitaries $U_{\mathrm{COE}}$ (top) and $U_{\mathrm{GOE}}$ (bottom) for $s = 1/2$. We choose $L = 8$ and average over $5000$ samples. The blue solid lines indicate the free probability result, \emph{i.e.}, the arcsine distribution \eqref{arcsinedist}. Both unitaries, although they exhibit different level statistics (left and right in Fig.\,\ref{rratiohaar}), drive the operators $S_1$ and $S_L$ to asymptotically free, so the plots become indistinguishable from each other.
} \label{haar2}
\end{figure}

Next, we perform the operator evolution with all unitaries in \eqref{unitaries}. Interestingly, we find that all of them drive the operators to asymptotic freeness. Figure \ref{haar2} shows the comparison between the Haar random ensembles and the exponential of the Gaussian ensembles. For simplicity, we only show the results for  $s = 1/2$, but $s = 1$ and $s = 3/2$ can be obtained straightforwardly. This illustrates that the asymptotic freeness is a more generic property and need not be tied to the Haar randomness of the unitary operator in the Hamiltonian evolution. It will be worthwhile to examine what properties of the unitary are responsible for freeness.

\section{Role of eigenvectors in freeness and decorrelated ensembles}
\label{app:DecorrEns}

As discussed in Sec.\,\ref{sec:operator_stat}, the eigenvalue statistics of operators of the form $S_{1}+S_{L}(t)$, with time evolution governed by a Hamiltonian drawn from the RP ensemble, coincide with the predictions of free probability only in the ergodic phase, where eigenvectors are fully delocalized. In contrast, in the fractal phase, although eigenvalue correlations persist, the corresponding statistics do not approach the free-probability limit even at late times, reflecting the fractal character of the eigenstates. To isolate the respective contributions of eigenvector delocalization and eigenvalue correlations in the emergence of freeness, we consider a variant of App.\,\ref{app:ExpGauss}. Specifically, we analyze operator statistics under time evolution generated by Hamiltonians drawn from a decorrelated version of the RP ensemble, building on previous work \cite{Camargo:2025zxr} where such a situation was considered for classical Gaussian random matrix ensembles.  

In essence, the decorrelated ensemble is constructed from a given random-matrix ensemble such that a Hamiltonian $H$ drawn from it preserves the localized, delocalized, or fractal structure of the original eigenvectors, while eliminating any correlations among the eigenvalues \cite{Nandy:2024zcd, Magan:2024nkr}. Concretely, consider a realization of matrices $A$ and $B$ sampled according to Eq.\,\eqref{HamRP}, yielding an RP Hamiltonian $H$ with eigenvectors $\lvert n\rangle$ and eigenvalues $E_{n}$ (ordered in ascending $n$). The spectrum defines a discrete density of states (DOS) $\rho(E)$, whose ensemble average $\overline{\rho(E)}$ approaches the $\gamma$-dependent DOS in the thermodynamic limit $N\to\infty$ (see Fig.\,\ref{fig:DOSschematic}). While the true eigenvalues $E_{n}$ fluctuate around $\overline{\rho(E)}$ and exhibit $\gamma$-dependent correlations, one may instead draw $N$ independent samples ${E_{n}'}$ from $\overline{\rho(E)}$. These ``decorrelated'' eigenvalues reproduce the same average DOS but, by construction, lack the original correlations and follow Poissonian level statistics for any $\gamma$ \cite{Nandy:2024zcd}. A new Hamiltonian $H'$ can then be defined by combining the original eigenvectors $\lvert n\rangle$ with the decorrelated eigenvalues $E_{n}'$, ordered ascendingly. This procedure can be repeated for arbitrary realizations of the RP ensemble, since every such Hamiltonian is diagonalizable with real eigenvalues.

\begin{figure}[t]
    \centering
    \begin{tabular}{ccc}
        \includegraphics[width=0.45\textwidth]{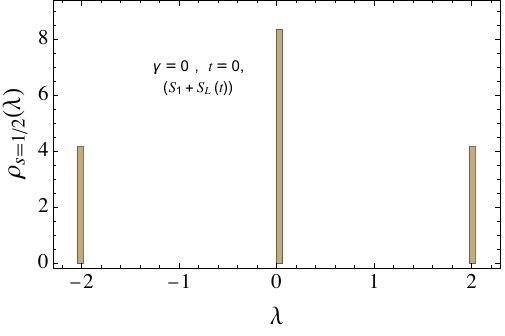} &
        \includegraphics[width=0.45\textwidth]{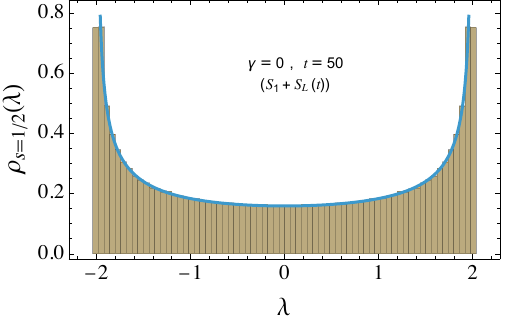} \\
       \multicolumn{2}{c}{\centering \includegraphics[width=0.45\textwidth]{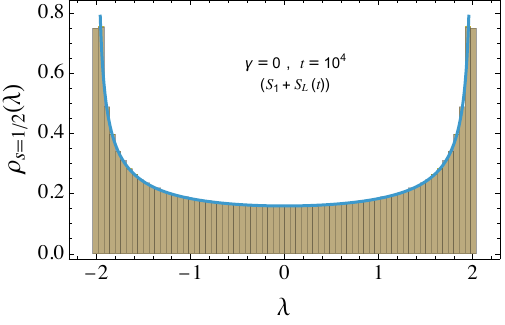}} \\
    \end{tabular}
    \caption{Statistics of eigenvalues of the sum of spin-$1/2$ operators $S_1(0)+S_{L}(t)$ for increasing values of time under time evolution by a Hamiltonian drawn from the RP ensemble in the ergodic phase $\gamma=0$ (yellow) and under time evolution by a Hamiltonian drawn from a decorrelated ensemble (gray). The overlap between these two distributions is shown in dark brown. The blue curve is the arcsine distribution. The bin size in each histogram is $\lbrace 0.06\rbrace$, $L=8$, and the number of independent realizations considered is $500$.}
    \label{fig:OpStatDecorrEnsemErgodic}
\end{figure}

\begin{figure}[t]
    \centering
    \begin{tabular}{ccc}
        \includegraphics[width=0.45\textwidth]{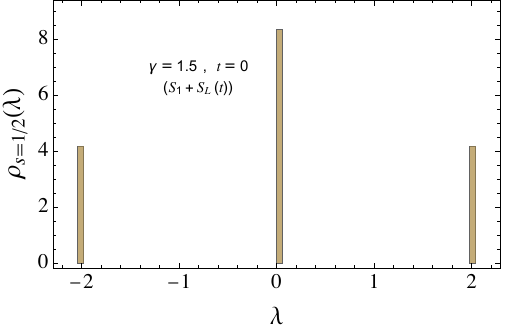} &
        \includegraphics[width=0.45\textwidth]{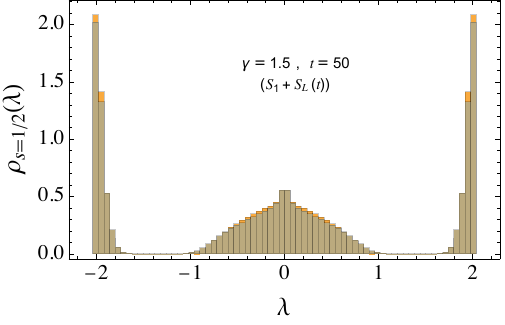} \\
       \multicolumn{2}{c}{\centering \includegraphics[width=0.45\textwidth]{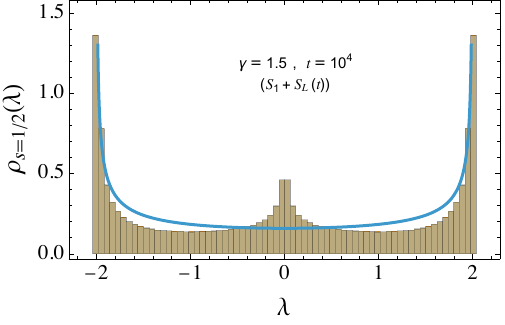}} \\
    \end{tabular}
    \caption{Statistics of eigenvalues of the sum of spin-$1/2$ operators $S_1(0)+S_{L}(t)$ for increasing values of time under time evolution by a Hamiltonian drawn from the RP ensemble in the fractal phase $\gamma=1.5$ (yellow) and under time evolution by a Hamiltonian drawn from a decorrelated ensemble (gray). The overlap between these two distributions is shown in dark brown. The blue curve is the arcsine distribution. The bin size in each histogram is $\lbrace 0.06\rbrace$, $L=8$, and the number of independent realizations considered is $500$.}
    \label{fig:OpStatDecorrEnsemFractal}
\end{figure}

\begin{figure}[t]
    \centering
    \begin{tabular}{ccc}
        \includegraphics[width=0.45\textwidth]{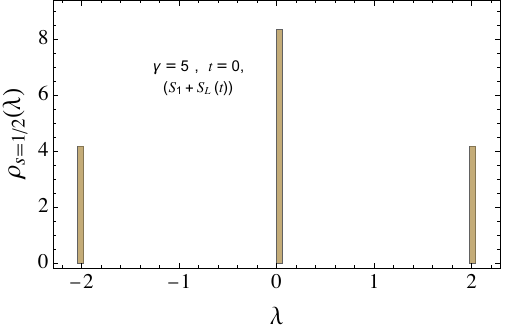} &
        \includegraphics[width=0.45\textwidth]{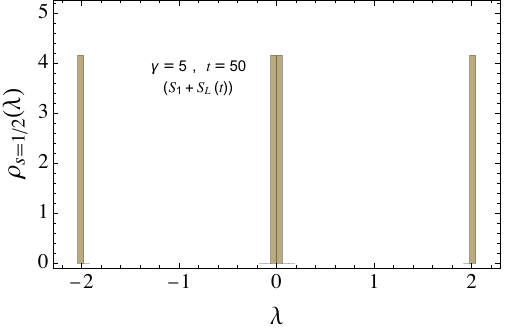} \\
       \multicolumn{2}{c}{\centering \includegraphics[width=0.45\textwidth]{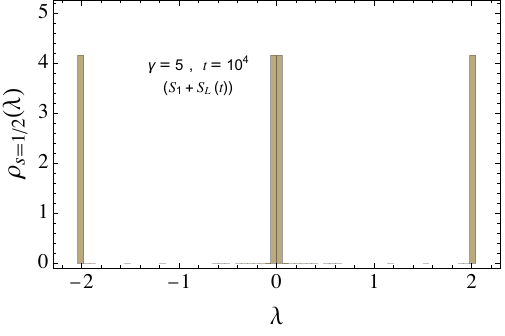}} \\
    \end{tabular}
    \caption{Statistics of eigenvalues of the sum of spin-$1/2$ operators $S_1(0)+S_{L}(t)$ for increasing values of time under time evolution by a Hamiltonian drawn from the RP class in the localized phase $\gamma=5$ (yellow) and under time evolution by a Hamiltonian drawn from a decorrelated ensemble (gray). The overlap between these two distributions is shown in dark brown. The bin size in each histogram is $\lbrace 0.06\rbrace$, $L=8$, and the number of independent realizations considered is $500$.}
    \label{fig:OpStatDecorrEnsemLocalized}
\end{figure}

Figures~\ref{fig:OpStatDecorrEnsemErgodic}-\ref{fig:OpStatDecorrEnsemLocalized} show the operator spectral statistics of the spin-$1/2$ operator sum $S_{1}+S_{L}(t)$ at different times for values of $\gamma$ corresponding to the ergodic ($\gamma=0$), fractal ($\gamma=1.5$), and localized ($\gamma=5$) phases, respectively. In the ergodic phase, the continuous DOS is the Wigner semicircle. For the fractal and localized regimes, we employ the corrected DOS obtained in Sec.\,\ref{sec:ApproxDOSforRP}, up to $O(\alpha^{2})$ with $\alpha=N^{-\gamma/2}$, as given in Eq.\,\eqref{GC_2}. This choice is appropriate since the values of $\gamma$ considered satisfy $\gamma>1$, consistent with the perturbative scheme discussed in Sec.\,\ref{sec:PertSchemeResolv}.

We observe that the operator spectral statistics generated by time evolution under decorrelated ensembles closely track those obtained from the full RP Hamiltonian, consistent with the results of Ref.\,\cite{Camargo:2025zxr} for Hamiltonians drawn from Hermitian Gaussian ensembles. Notably, this agreement holds even at intermediate timescales, before the statistics converge to their asymptotic distributions at late times. These results reinforce the conclusion of Ref.\,\cite{Camargo:2025zxr} that the eigenvectors play the central role in determining the operator statistics of sums of the form $X_{i}(0)+X_{j}(t)$, and that in the ergodic case they suffice to reproduce the free-probability prediction regardless of eigenvalue correlations. In the fractal phase, however, we observe small deviations at intermediate times, whereas in the localized phase, the decorrelated and true statistics remain nearly indistinguishable across all timescales.

\section{Free additive convolution of spin-$s$ operators} \label{app:SumSpinS}
In this appendix, we review the derivation of the free probability prediction for the spectrum of the sum of two free spin-$s$ operators, whose spectral distributions take the form:
\begin{equation} \label{eq:spectralSpinS}
\rho_{s}(\lambda)=\frac{1}{2s+1}\sum_{j=-s}^{s}\delta(\lambda-j)\,.
\end{equation}
The corresponding Cauchy transform is given by
\begin{equation} \label{eq:CauchySpinS}
G_{s}(z)= \int \frac{\rho_{s}(\lambda)}{z-\lambda}=\frac{1}{2s+1}\sum_{j=-s}^{s} \frac{1}{z-j}\,.
\end{equation}
The associated $R$-transform, defined as $R_s(z)=G^{-1}_{s}(z)-1/z$, is obtained by solving an algebraic equation of order $s$. This poses a practical challenge for large values of $s$: for example, computing the $R$-transform for spin-$3/2$ operators already involves solving a quartic equation, with even higher-degree equations required for larger spin values. It thus becomes evident that one must resort to numerical methods in such cases. 

A particularly effective numerical technique for evaluating the free additive convolution of random variables with arbitrary measures is the subordination function method. We refer to Sec.\,5 of \cite{speicher2019lecturenotesfreeprobability} for a thorough discussion of this method, including detailed derivations of all formulas used here.

Suppose we are given two free random variables $a$ and $b$, with Cauchy transforms $G_a(z)$ and $G_b(z)$, respectively. It can be shown that the Cauchy transform of their sum $\rho_a \boxplus \rho_b$ is given by~\cite{speicher2019lecturenotesfreeprobability}
\begin{equation} \label{eq:numCauchySum}
G_{\rho_a \boxplus \rho_b}(z) = G_a(\omega_a(z))\,,
\end{equation}
where the {\it subordination function} $\omega_a(z)$ satisfies the fixed-point equation for $\omega_a(z) \in \mathbb{C}^+$ (the upper half-plane):
\begin{equation} \label{eq:iteration}
\omega_a(z) = z + H_b\left[ z + H_a(\omega_a(z)) \right]\,,
\end{equation}
with
\begin{equation}
H_a(z) = \frac{1}{G_a(z)} - z\,,~~~~ H_b(z) = \frac{1}{G_b(z)} - z\,,
\end{equation}
being auxiliary functions. Finally, the spectral distribution of $a + b$ is obtained via the inverse Stieltjes transform of \eqref{eq:numCauchySum}:
\begin{equation} \label{eq:inverseNumSum}
\rho_{a+b}(\lambda) = -\lim_{\epsilon \rightarrow 0} \frac{1}{\pi} \text{Im}\, G_{\rho_a \boxplus \rho_b}(\lambda + i \epsilon)\,.
\end{equation}

Now we illustrate the application of this method by using it to compute the spectral density of the sum of two spin operators with $s = 10$. This value is chosen to demonstrate that the method remains applicable for relatively large $s$, although the resulting expressions become increasingly complicated as $s$ increases.

\begin{figure}
    \centering
\includegraphics[width=0.6\linewidth]{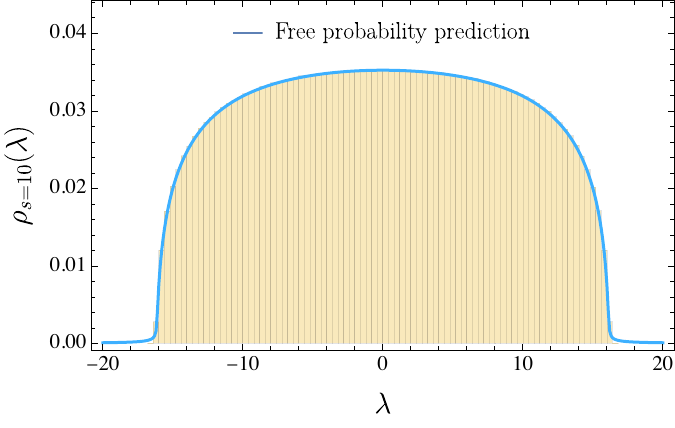}
    \caption{Spectral distribution of the sum of two spin-10 operators, $a = S_{1}$ and $b = U^\dagger S_{1} U$, where $U$ is a Haar-random unitary and $S_1$ is defined in Eq.~\eqref{bop} with $s = 10$ and $L = 2$. The solid line shows the free probability prediction, compared with histogram data obtained from the direct diagonalization of $a + b$ over $10^4$ realizations.}
    \label{fig:SumSpin10}
\end{figure}

\paragraph{Example: Sum of two free spin-10 operators:} Consider the sum of two free spin-10 operators, $a = S_{1}$ and $b = U^\dagger S_{1} U$, where $U$ is a Haar-random unitary and $S_1$ is defined in Eq.\eqref{bop} for $s=10$ and $L=2$. Both operators share the same spectral distribution and Cauchy transform, given by \eqref{eq:spectralSpinS} and \eqref{eq:CauchySpinS} for $s = 10$. The corresponding auxiliary functions are:
\begin{equation} \label{eq:Hspin10}
H_a(z) = H_b(z) = \frac{P(z)}{Q(z)}\,.
\end{equation}
where $P(z)$ and $Q(z)$ are given by
\begin{align}
P(z) &= 2z \Big(
 \big(
-41309657025 z^6 
+ 911087234376 z^4 
- 11156038265680 z^2 \big) z^4 \notag \\
&\quad 
-183668715648000 z^2 
+- 11 \big( 
35 z^6 
- 11172 z^4 
+ 1443810 z^2 
- 97774684 
\big) z^{12} \notag \\
&\quad 
+ 69514526366208 z^4 
+ 131681894400000 
\Big)\,, \label{eq:Pz}
\end{align}
\begin{align}
Q(z) &= 21 z^{20} 
- 7315 z^{18} 
+ 1044582 z^{16} 
- 79409550 z^{14} + 3495444953 z^{12} \notag \\
&\quad  
- 90881245455 z^{10} 
+ 1366630851564 z^8 - 11156038265680 z^6 \notag \\
&\quad 
+ 43446578978880 z^4 
- 61222905216000 z^2 
+ 13168189440000\,. \label{eq:Qz}
\end{align}
After numerically solving \eqref{eq:iteration} for $H_a$ and $H_b$ given by \eqref{eq:Hspin10}, and using \eqref{eq:inverseNumSum}, we obtain the spectral density $\rho_{a+b}(\lambda)$ numerically. The result is shown in Fig.\,\ref{fig:SumSpin10} and compared to the numerical data obtained by direct diagonalization of $a + b$.

\section{Operator Dependence of Freeness}\label{app:op_dependence}

\begin{figure}
       \hspace*{-0.4 cm}
    \includegraphics[width=1.03\linewidth]{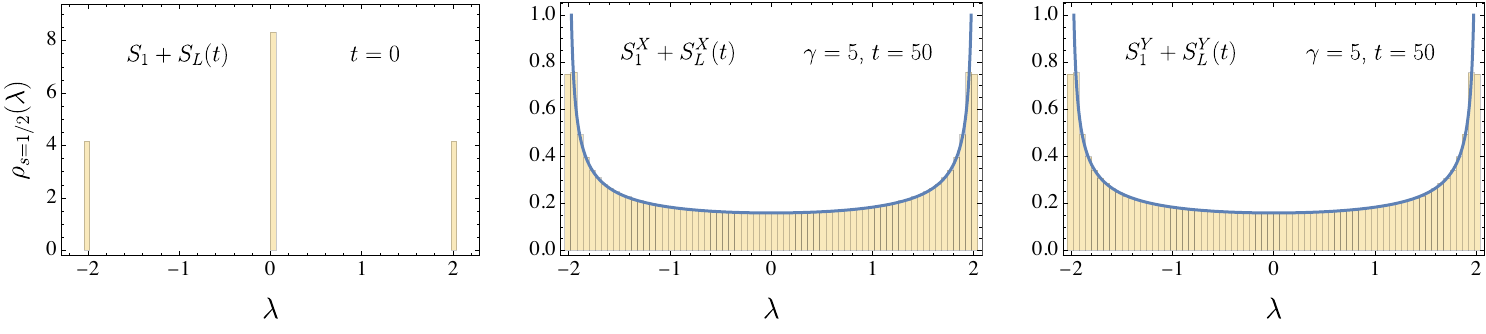}
    \caption{The statistics of the eigenvalues of the operators defined in Eqs.,\eqref{aop}–\eqref{bop}, with the $Z$ operator replaced by $X$ or $Y$ for $s=1/2$. Results are shown at the initial time $t=0$ (left), and at a later time $t=50$ under evolution with the $X$ operator (middle) and the $Y$ operator (right). We set $L=8$ and average over 5000 independent Hamiltonian realizations in the localized phase ($\gamma=5$). The solid blue line denotes the arcsine distribution predicted by free probability (Eq.\,\eqref{arcsinedist}).}
    \label{fig:OperatorStatisticsX}
\end{figure}

In this appendix, we examine the extent to which our results depend on the choice of operators. In the main text, we focused on diagonal operators inspired by generalized $Z$-Pauli matrices corresponding to spin-$s$ representations of $SU(2)$. This choice has special consequences in the context of the RP model, where the Hamiltonian consists of a diagonal matrix plus a random matrix suppressed by a factor of $N^{\gamma/2}$. For $\gamma > 1$ and sufficiently large $N$, the off-diagonal part becomes negligible, rendering the diagonal operators approximately conserved and quasi-local.

We demonstrated that the spectral statistics predicted by free probability theory emerge in the ergodic phase for these diagonal operators. However, such behavior does not extend to the fractal or localized phases, where the spectra deviate significantly from free probability predictions. This raises the natural question of whether different classes of operators---particularly non-diagonal ones such as those constructed by $X$- or $Y$-Pauli matrices--might lead to different conclusions. Unlike the diagonal case, these operators are not approximately conserved in the RP model. It is therefore conceivable that the spectral statistics associated with free probability might emerge at late times even in the fractal or localized phases, provided one focuses on such non-diagonal operators.

\begin{figure}
    \centering
    \includegraphics[width=1\linewidth]{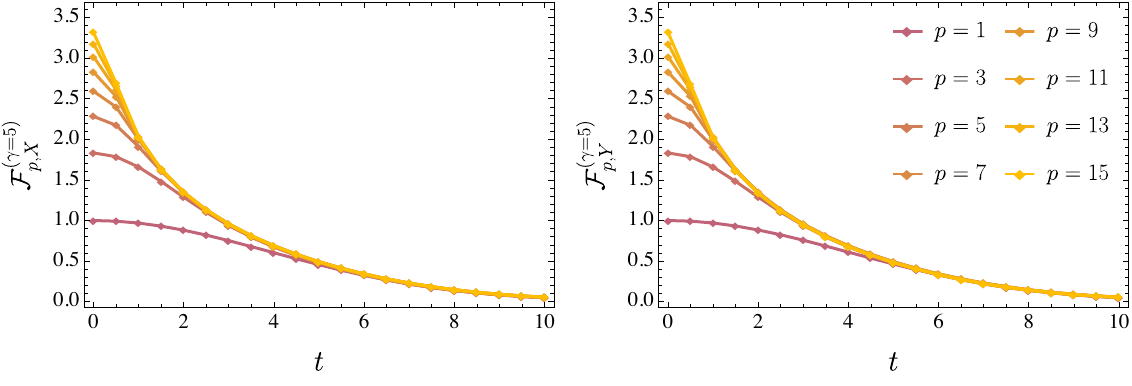}
    \caption{Cumulative OTOCs for $X$ (left) and $Y$ (right) operators in the localized phase for the operator at site $i=1$ in the $s=1/2$ case. Here, $p$ denotes the number of terms included in the cumulative OTOC \eqref{sumosquared}. We choose $f(n) = 1/n$, consider a system size of $L=8$, and average over 50 independent realizations of the Hamiltonian. The timescale can be compared with Fig.\,\ref{fig:plotallgamma} to assess when the cumulative OTOC vanishes.}
    \label{fig:cumOTOCXY}
\end{figure}

Figure \ref{fig:OperatorStatisticsX} shows the eigenvalue density of $S_1^X(0)+S_L^X(t)$ (middle) and $S_1^Y(0)+S_L^Y(t)$ (right) for spin-$1/2$ operators at $t=0$ and $t=50$. The superscript denotes the chosen operator, \emph{i.e.}, by replacing the $Z$ operators in Eqs.\,\eqref{aop}-\eqref{bop} by $X$ and $Y$ operators respectively. In both cases, the distribution converges to the free probability prediction, \emph{i.e.}, the arcsine law (Eq.\,\eqref{arcsinedist}) at sufficiently late times, despite the system being in the localized phase ($\gamma=5$). This observation is corroborated by the vanishing cumulative OTOCs at late times (see Fig.\,\ref{fig:cumOTOCXY}), although the associated timescale differs from that of the $Z$ operators.

The above results indicate that freeness between certain pairs of operators can emerge even in the absence of spectral chaos and despite localized eigenvectors. However, this emergence is operator-dependent and fails to occur for approximately conserved local charges (e.g., $Z$-operators, which are diagonal in the computational basis). We also showed that the mechanism underlying the asymptotic freeness of $X$- and $Y$-operators in the localized phase differs from that in the ergodic phase. In the ergodic phase, delocalized eigenvectors likely generate asymptotic freeness for essentially all operators, whereas in the localized phase, the effect arises only for specific operators and is driven by the random phases these operators acquire under time evolution. Thus, with carefully chosen operators, operator statistics may serve as a useful tool to classify localized phases in many-body systems. We plan to investigate this further, including tests of whether the observed freeness originates from delocalized eigenvectors or from a distinct mechanism.

\bibliographystyle{JHEP}
\bibliography{references}  
\end{document}